\documentclass{nature}
\usepackage{aas_macros}
\usepackage{graphicx}
\usepackage{pdflscape}
\usepackage{hyperref}
\usepackage{txfonts}
\usepackage{enumitem}
\usepackage{textcomp}
\usepackage{threeparttable}
\usepackage{xcolor}
\usepackage{booktabs}
\usepackage[normalem]{ulem}
\usepackage{lineno}
\usepackage{pdfpages}
\usepackage{subfigure}
\usepackage{comment}
\makeatletter
\usepackage[symbol]{footmisc}

\usepackage{longtable}
\usepackage{dutchcal}
\usepackage{marvosym}
\usepackage{lipsum}
\usepackage{float}
\usepackage[utf8]{inputenc}
\usepackage[T1]{fontenc}

\let\saved@includegraphics\includegraphics
\AtBeginDocument{\let\includegraphics\saved@includegraphics}

\makeatother

\newcommand{\arcdeg}{\mbox{$^{\circ}$}}
\newcommand{\arcmin}{\mbox{$^{\prime}$}}
\newcommand{\arcsec}{\mbox{$^{\prime\prime}$}}
\newcommand{\farcs}{\mbox{$.\!\!^{\prime\prime}$}}

\newcommand{\dmu}{~pc~cm$^{-3}$}
\newcommand{\frb}{FRB 20190520B}
\newcommand{\realf}{\textit{realfast}}

\def\DM{\mathrm{DM}}
\def\DMm{\DM_{\rm MW}}
\def\DMmwh{\DM_{\rm MW,halo}}

\def\DMi{\DM_{\rm IGM}}
\def\DMh{\DM_{\rm host}}

\newcommand{\DMunits}{{\rm pc~cm^{-3}}}

\def\Ha{\mathrm{H}\alpha}

\definecolor{dkblue}{RGB}{54, 86, 169}
\definecolor{red}{RGB}{200, 10, 10}

\title{A repeating fast radio burst associated with a persistent radio source}
\author{
C.-H. Niu$^{1}$\href{https://orcid.org/0000-0001-6651-7799}{\includegraphics[scale=0.08]{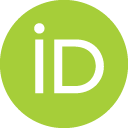}}\footnotemark[1] ,
K. Aggarwal$^{2,6}$\href{https://orcid.org/0000-0002-2059-0525}{\includegraphics[scale=0.08]{ORCIDiD.png}}\footnotemark[1] ,
\textsuperscript{\Letter}D. Li$^{1,12}$ \href{https://orcid.org/0000-0003-3010-7661}{\includegraphics[scale=0.08]{ORCIDiD.png}\footnotemark[1] }, 
X. Zhang$^{3,12}$\href{https://orcid.org/0000-0002-8086-4049}{\includegraphics[scale=0.08]{ORCIDiD.png}},
S. Chatterjee$^{4}$\href{https://orcid.org/0000-0002-2878-1502}{\includegraphics[scale=0.08]{ORCIDiD.png}},
C.-W. Tsai$^{1}$\href{https://orcid.org/0000-0002-9390-9672}{\includegraphics[scale=0.08]{ORCIDiD.png}},
\textsuperscript{\Letter}W. Yu$^{3}$\href{https://orcid.org/0000-0002-3844-9677}{\includegraphics[scale=0.08]{ORCIDiD.png}}, 
\textsuperscript{\Letter}C. J.~Law$^{5,8}$\href{https://orcid.org/0000-0002-4119-9963}{\includegraphics[scale=0.08]{ORCIDiD.png}, }
S. Burke-Spolaor$^{2,6,7}$\href{https://orcid.org/0000-0003-4052-7838
}{\includegraphics[scale=0.08]{ORCIDiD.png}}, 
J. M. Cordes$^{4}$\href{https://orcid.org/0000-0002-4049-1882}{\includegraphics[scale=0.08]{ORCIDiD.png}}, 
Y.-K. Zhang$^{1,12}$\href{https://orcid.org/0000-0002-8744-3546}{\includegraphics[scale=0.08]{ORCIDiD.png}},
S. K. Ocker$^{4}$\href{https://orcid.org/0000-0002-4941-5333}{\includegraphics[scale=0.08]{ORCIDiD.png}}, 
J.-M. Yao$^{13}$,
P. Wang$^{1}$, 
Y. Feng$^{1,9,12}$\href{https://orcid.org/0000-0002-0475-7479}{\includegraphics[scale=0.08]{ORCIDiD.png}},
Y. Niino$^{10,20}$\href{https://orcid.org/0000-0001-5322-5076}{\includegraphics[scale=0.08]{ORCIDiD.png}}, 
C. Bochenek$^{5}$,
M. Cruces$^{17}$\href{https://orcid.org/0000-0001-6804-6513}{\includegraphics[scale=0.08]{ORCIDiD.png}} ,
L. Connor$^{5}$,
J.-A. Jiang$^{18}$\href{https://orcid.org/0000-0002-9092-0593}{\includegraphics[scale=0.08]{ORCIDiD.png}}, 
S. Dai$^{14,19}$, 
R. Luo$^{14}$\href{https://orcid.org/0000-0002-4300-121X}{\includegraphics[scale=0.08]{ORCIDiD.png}}, 
G.-D. Li$^{1,12}$\href{https://orcid.org/0000-0003-4007-5771}{\includegraphics[scale=0.08]{ORCIDiD.png}},
C.-C. Miao$^{1,12}$,
J.-R. Niu$^{1,12}$, 
R. Anna-Thomas$^{2,6}$\href{https://orcid.org/0000-0001-8057-0633}{\includegraphics[scale=0.08]{ORCIDiD.png}},
J. Sydnor$^{2,6}$\href{https://orcid.org/0000-0002-3360-9299}{\includegraphics[scale=0.08]{ORCIDiD.png}}, 
D. Stern$^{11}$\href{https://orcid.org/0000-0003-2686-9241}{\includegraphics[scale=0.08]{ORCIDiD.png}},
W.-Y. Wang$^{1,15}$\href{https://orcid.org/0000-0001-9036-8543}{\includegraphics[scale=0.08]{ORCIDiD.png}}, 
M. Yuan$^{1,12}$, 
Y.-L. Yue$^{1}$, 
D.-J. Zhou$^{1,12}$,
Z. Yan$^{3}$,
W.-W. Zhu$^{1}$, 
B. Zhang$^{16}$\href{https://orcid.org/0000-0002-9725-2524}{\includegraphics[scale=0.08]{ORCIDiD.png}}
}

\spacing{1.2}

\begin{document}
\maketitle
\footnotetext[1]{These authors contributed equally to this work.}

\begin{affiliations}
\item National Astronomical Observatories, Chinese Academy of Sciences, Beijing 100101, China
\item Department of Physics and Astronomy, West Virginia University, P.O. Box 6315, Morgantown, WV 26506, USA
\item Shanghai Astronomical Observatory, Chinese Academy of Sciences, 80 Nandan Road, Shanghai 200030, China
\item Cornell Center for Astrophysics and Planetary Science, and Department of Astronomy, Cornell University, Ithaca, NY 14853, USA
\item Cahill Center for Astronomy and Astrophysics, MC 249-17 California Institute of Technology, Pasadena, CA 91125, USA
\item Center for Gravitational Waves and Cosmology, West Virginia University, Chestnut Ridge Research Building, Morgantown, WV 26505, USA
\item Canadian Institute for Advanced Research, CIFAR Azrieli Global Scholar, MaRS Centre West Tower, 661 University Ave. Suite 505, Toronto ON M5G 1M1, Canada
\item Owens Valley Radio Observatory, California Institute of Technology, 100 Leighton Lane, Big Pine, CA, 93513, USA
\item Research Center for Intelligent Computing Platforms, Zhejiang Laboratory, Hangzhou 311100, China
\item Institute of Astronomy, Graduate School of Science, The University of Tokyo, 2-21-1 Osawa, Mitaka, Tokyo 181-0015, Japan
\item Jet Propulsion Laboratory, California Institute of Technology, 4800 Oak Grove Drive, M/S 264-723, Pasadena, CA 91109, USA
\item University of Chinese Academy of Sciences, Beijing 100049, China
\item Xinjiang Astronomical Observatory, Chinese Academy of Sciences, 150, Science 1-Street, Urumqi, Xinjiang 830011, China
\item CSIRO Space and Astronomy, PO Box 76, Epping, NSW 1710, Australia
\item Department of Astronomy, School of Physics, Peking University, Beijing 100871, China
\item Department of Physics and Astronomy, University of Nevada, Las Vegas, Las Vegas, NV 89154, USA
\item Max-Planck-Institut für Radioastronomie, Auf dem Hügel 69, D-53121 Bonn, Germany
\item Kavli Institute for the Physics and Mathematics of the Universe (WPI), The University of Tokyo, 5-1-5 Kashiwanoha, Kashiwa, Chiba 277-8583, Japan
\item School of Science, Western Sydney University, Locked Bag 1797, Penrith South DC, NSW 2751, Australia
\item Research center for the early universe, The University of Tokyo, Hongo, 7-3-1, Bunkyo-ku, Tokyo, 113-0033, Japan 
\end{affiliations}

\begin{abstract}
The dispersive sweep of fast radio bursts (FRBs) has been used to probe the ionized baryon content of the intergalactic medium\cite{Macquart_2020}, which is assumed to dominate the total extragalactic dispersion.
While the host galaxy contributions to dispersion measure (DM) appear to be small for most FRBs\cite{2020arXiv200512891N}, in at least one case there is evidence for an extreme magneto-ionic local environment\cite{mic18,hi21} and a compact persistent radio source\cite{Cha17}.
Here we report the detection and localization of the repeating FRB 20190520B, which is co-located with a compact, persistent radio source and associated with a dwarf host galaxy of high specific star formation rate at a redshift \textbf{$z=0.241\pm0.001$}. The estimated host galaxy DM~$\approx 903^{+72}_{-111}$~pc~cm$^{-3}$, nearly an order of magnitude higher than the average of FRB host galaxies\cite{2018MNRAS.481.2320L,2020arXiv200512891N}, far exceeds the DM contribution of the intergalactic medium. Caution is thus warranted in inferring redshifts for FRBs without accurate host galaxy identifications. 

\end{abstract}

FRB 20190520B was detected with the Five-hundred-meter Aperture Spherical radio Telescope (FAST)\cite{NAN_2011} in drift-scan mode as part of the Commensal Radio Astronomy FAST Survey (CRAFTS)\cite{Li_2019} at 1.05--1.45~GHz in 2019. Four bursts were detected during the initial 24s scan. Monthly follow-up tracking observations between 2020 April and 2020 September detected 75 bursts in 18.5~hrs with a mean pulse DM of $1204.7\pm 4.0$~\dmu. Assuming a Weibull distribution of the burst waiting time, we model the FRB burst rate to be $R = 4.5^{+1.9}_{-1.5}\ \rm hr^{-1}$, for a fluence lower limit of 9 mJy ms and a burst width of 1 ms, indicating that the FRB can episodically have a high burst rate.
Similar to other repeating FRBs, this FRB shows complex frequency-time intensity structure\cite{Hessels2019} with multi-component-profiles, sub-burst drifting, and scattering (\ref{fig:Water-Fall plots} and Methods). 
We detected no linear polarization for \frb~from FAST observation (see Methods), the rotation measure (RM) for this source have been detected from higher frequency band \cite{Feng22,Reshma22}. The properties of \frb~can be found in Table \ref{tab:generaltab}.

We localized \frb\ with the Karl G. Jansky Very Large Array (VLA) using the \realf\ fast transient detection system\cite{law18}. Throughout the second half of 2020, we observed the source for 16 hours and detected 3, 5, and 1 bursts in bands centered at 1.5~GHz, 3~GHz, and 5.5~GHz, respectively. We measured a burst source position in the International Celestial Reference Frame (ICRF) of (R.A., Decl.)~[J2000] = (16h02m04.272s, $-11$\arcdeg17\arcmin17.32\arcsec) with positional uncertainty (1$\sigma$) of (0.10\arcsec, 0.08\arcsec) dominated by systematic effects (see Methods). Deep images using data from the same observing campaign revealed a persistent radio continuum counterpart at (R.A., Decl.) [J2000] = (16h02m04.261s, $-11$\arcdeg17\arcmin17.35\arcsec) with 1$\sigma$ position uncertainty of (0.10\arcsec, 0.05\arcsec)  that is compact ($<$0.36\arcsec; see Methods) and has a flux density of $202~\pm~8~\mu$Jy averaged over span of $\sim$~2 months from August 30 to November 16 of 2020 at 3.0 GHz.
Using the average flux density of each sub-band over the VLA campaign, we find that the radio continuum counterpart spectrum can be fit with a power-law spectral index of $-0.41\pm$0.04 (see Methods). 

In order to identify the host galaxy, an optical image ($R^{\prime}$-band) was obtained using the Canada-France-Hawaii Telescope/MegaCam. \ref{fig:sky} shows the FRB location compared to deep optical, near-infrared (NIR), and radio images of the field. The optical image ($R^{\prime}$-band) reveals the galaxy J160204.31$-$111718.5 at the location of the FRB, the light profile of which peaks at $\sim 1$\arcsec\ south east. Given the measured offset of 1.3\arcsec\ between the FRB and the peak of the galaxy light profile, and the sky surface density of galaxies with this magnitude, we estimate a chance coincidence probability of 0.8\% (see Methods), indicating that J160204.31$-$111718.5 is the host galaxy of \frb. The NIR image ($J$-band, 1.153$-$1.354 $\mu$m) obtained with Subaru/MOIRCS\cite{2008PASJ...60.1347S} most likely shows the stellar continuum emission of the host galaxy with an AB magnitude of 22.1$\pm$0.1 in $J$-band, with \frb\ and the radio continuum counterpart on the galaxy periphery. 

We obtained an optical spectrum at the location of the FRB with the Double Spectrograph on the Palomar 200-inch Hale Telescope, which reveals the redshift of the putative host to be $z = 0.241\pm0.001$ based on a detection of strong H$\alpha$, [O\,III] 4859\AA, and [O\,III] 5007\AA\ lines (see Methods). A follow-up observation with the Low Resolution Imaging Spectrometer (LRIS) at the Keck I Telescope covering both the FRB location and the nearby Subaru $J$-band source along the extended $R^{\prime}$-band structure indicates that the $R^{\prime}$-band structure is dominated by the [O III] emission at the same redshift of $z=0.241$. The H$\alpha$ luminosity $L_{H\alpha} = 7.4\pm0.2 \times 10^{40}$ erg s$^{-1}$ after extinction correction implies a star formation rate of $\sim$ 0.41 M$_{\odot}$ yr$^{-1}$. Based on the $J$-band magnitude, we estimate the stellar mass of the host galaxy to be $\sim 6 \times 10^8$ M$_{\odot}$. Thus, we characterize J160204.31$-$111718.5 as a dwarf galaxy with a relatively high star-formation rate for its stellar mass compared with local SDSS galaxies\cite{2004MNRAS.351.1151B}. 
At the luminosity distance of 1218 Mpc implied by the redshift, the radio continuum counterpart has a spectral luminosity of $L_{\rm{3~GHz}}=3\times10^{29}$\ erg s$^{-1}$\ Hz$^{-1}$. 

The redshift of an FRB source is often estimated from the DM attributed to the intergalactic medium (IGM), $\rm DM_{IGM} = DM_{FRB}-DM_{host}-DM_{MW}$. Theoretical calculations\cite{2003ApJ...598L..79I} and observations\cite{Macquart_2020} have independently estimated the IGM contribution to FRB DM as a function of redshift (\ref{fig:scatter_dm_z}). For a DM contribution from the Milky Way of
$113\pm17$ \dmu~(including a $\pm $40\% range for the NE2001 estimate\cite{Cordes03} of the disk contribution and a uniform halo distribution from 25 to 80~\dmu) combined with an assumed host galaxy DM of only 50 \dmu, the implied redshift range for \frb\ is $z \sim 2.2$ to 0.9 for baryon fractions $f_{\rm IGM}$ of 0.6 to 1 for the ionized IGM, much larger than the measured value (see Methods).

Instead, using $\rm DM_{IGM}(z=0.241)
\simeq 195^{+110}_{-70}$~\dmu\
(68\% interval to account for
 cosmic variance of $\rm DM_{IGM}$ and 0.85 for the ionized baryon fraction\cite{2018ApJ...867L..21Z}) combined with the Milky Way contribution,
 a very large host DM (disk + circumgalaxy + local to source contribution)
of $\DMh = 903^{+72}_{-111}~\DMunits$ is inferred from the posterior distribution. Over a broader range of ionized fractions from 0.6 to 1, $\DMh$ extends from 1020 to 745~\dmu.

In addition to the low chance coincidence probability, the measured DM and scattering properties of \frb\  
render it unlikely that J160204.31$-$111718.5 is a foreground galaxy with the true FRB host much further in the background. 
The observed H$\alpha$ emission implies a DM contribution from warm gas in the galaxy ranging from $230$ to $650$
pc cm$^{-3}$ (observer frame) for temperatures from $0.5\times10^4$ to $5 \times 10^4$ K (see Methods). The scattering contribution of a foreground galaxy depends not only on its DM contribution but also on the geometric leverage effect, which will increase the scattering by several orders of magnitude relative to scattering in the host galaxy. If J160204.31$-$111718.5 were a foreground galaxy, we estimate a fiducial range of scattering times 
from 0.6 to 20 s at 1.25 GHz, orders of magnitude larger than the observed mean scattering time of $10 \pm 2$ ms.
While this estimate depends on parameters such as the ionized gas temperature, path length through the galaxy, and the degree of turbulence in the gas, which are not well constrained, the observed scattering time is likely too small for the proposed host galaxy to lie in the foreground (see Methods).

The large $\DMh$ inferred for \frb~demonstrates that the distribution of ${\rm DM_{host}}$ values for the FRB population has a long tail, which may add considerable variance to estimates for the IGM contribution. It is conceivable that the ${\rm DM_{host}}$ distribution may differ for repeating and non-repeating FRBs, which could make non-repeating FRB DMs more accurate proxies for redshift.
To understand this further, a larger sample of precisely localized FRBs with measured host galaxy redshifts is needed to statistically characterize host galaxies, their circumgalactic media, and near-source environments along with the IGM\cite{2021ApJ...917...75M}. Accordingly, searches need to accommodate large values of $\DMh$ as part of the DM budget for FRB sources. Large $\DMh$ contributions may also imply large scattering, which can also reduce search sensitivity. 

The co-located radio continuum counterpart, the star-forming dwarf host galaxy, and the high repetition rate make \frb~ a clear analog to FRB 20121102A, the first known
repeating FRB\cite{Spi16} and the first to be identified with a compact, luminous persistent radio source (PRS)\cite{Cha17, 2017ApJ...834L...8M} ($L_{\rm 1.6\ GHz}\approx3\times10^{29}$~erg s$^{-1}$~Hz$^{-1}$). Another repeating source, FRB 20201124A, was also associated with a radio continuum counterpart \cite{2021ATel14497....1C, 2021ATel14515....1D, 2021ATel14549....1R}; however, optical spectroscopy and radio interferometric measurements demonstrated that the persistent radio emission was spatially extended and consistent with an origin in star formation in the host galaxy\cite{2021arXiv210609710R}. By contrast, the continuum counterpart to \frb\ appears to not be from star formation because its luminosity would imply a star-formation rate of $\sim$10~M$_{\odot}$~yr$^{-1}$, a factor of 25 larger than that measured for the host galaxy and a factor of five larger than the highest observed star formation rate for galaxies of this mass \cite{2004MNRAS.351.1151B}.
Given its extreme luminosity, unresolved structure in VLA observations, and offset from the peak optical emission in the host galaxy, we conclude that the radio continuum counterpart is a compact source ($<1.4\ \rm kpc$ ) physically connected to \frb, a PRS like that associated with FRB 20121102A (which has a VLBI-confirmed size $<0.7$ pc\cite{2017ApJ...834L...8M}). 

More than a dozen FRBs were localized prior to \frb, including five repeating sources\cite{2020ApJ...903..152H, 2021ApJ...910L..18B}, but only FRB 20121102A had been associated with a compact PRS. FRB 20121102A also demonstrates a sporadically large burst rate\cite{2021Natur.598..267L} (e.g.~a peak burst rate of 122 hr$^{-1}$), a substantial rotation measure that varies over both short (burst to burst) and long (year) timescales\cite{mic18,hi21}, and $\DMh$ as large as $\sim$300\dmu, suggesting that burst activity may be correlated with both relativistic plasma emitting synchrotron radiation and the presence of thermal plasma in the local FRB environment. In addition to a PRS, \frb\ shows a $\DMh$ that is almost three times larger than that of FRB 20121102A, and may have a comparably large RM (Methods). If a sizeable fraction of $\DMh$ is from thermal plasma in the circum-source medium, then perhaps the presence of a PRS and a dynamic magneto-ionic environment are correlated with FRB formation, and repeating FRBs like FRB 20121102A and \frb\ are younger sources that still reside in their complex natal environments. 

However, until further VLBI observation constraints become available, it is unclear how much of the large $\DMh$ for \frb\ is attributable to ionized gas in the circum-source medium vs. the host galaxy's interstellar medium.
Moreover, other repeating FRBs have very deep limits on PRS counterparts\cite{2020Natur.577..190M, 2021arXiv210511445K}, which complicates the connection between burst and magneto-ionic activity and PRS luminosity. For example, FRB 20200120E has an upper limit on persistent radio luminosity $L_{\rm 1.5\ GHz} < 3.1\times10^{23}$ erg s$^{-1}$ Hz$^{-1}$ \cite{2021arXiv210511445K}. It is associated with an old stellar population and has a modest $\DMh\lesssim50$ \dmu~where are in sharp contrast with FRB 20121102A and \frb. One possibility is that there are multiple kinds of sources that can emit FRBs, a point that has been argued based on burst rate and phenomenology\cite{2021ApJ...923....1P,2022MNRAS.510.1867L}. Alternatively, the source properties may evolve as the source ages, the PRS fades, the event rate drops, and the surrounding plasma dissipates\cite{2018ApJ...861..150P}.The similarity of \frb\ to FRB 20121102A suggests a potential connection between burst activity, the presence of a PRS, and $\DMh$ for at least some FRBs.

Various methods have been used to argue that repeating and non-repeating FRBs comprise different subclasses\cite{2021ApJ...923....1P,Hessels2019,mic18}, either at different evolutionary phases or entirely different physical scenarios. While the observed burst repetition and morphology can be time-dependent due to various mechanisms\cite{2020MNRAS.497.3076C},
PRS emission and ${\rm DM_{host}}$ may reflect more persistent aspects of the FRB environment and thus may be more reliable tracers of any putative subclasses.  Further progress will result from better characterization of the full range of host galaxy DMs, which will also help mitigate biases in the DM-redshift relation\cite{Macquart_2020}.

\clearpage

\begin{figure*}
\centering
\includegraphics[width=1\textwidth]{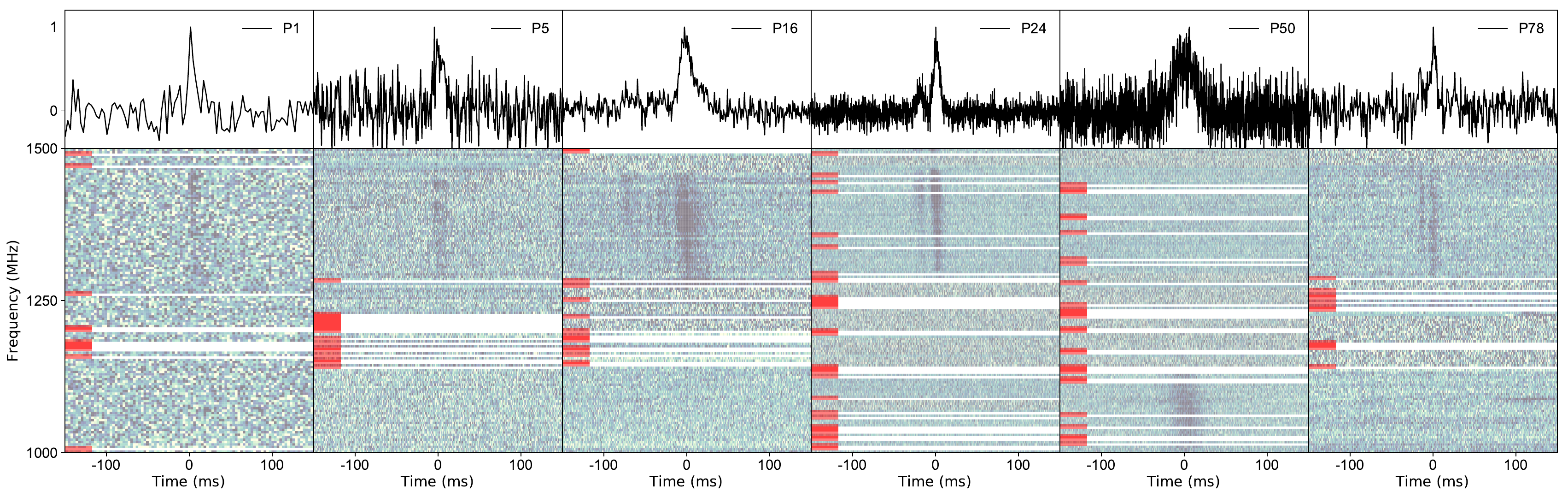}
\caption{
\footnotesize \textbf{Bursts from FRB 20190520B, shown as dynamic spectra and burst profiles.} Top panel: Frequency integrated burst normalized intensities that were detected during FAST observations. These six bursts are chosen from different observation epochs. The burst labels `PXX' are in order of arrival time and the corresponding burst properties are shown in supplementary information Table 1.
Bottom panel:  De-dispersed dynamic spectra of the bursts clearly showing their band-limited nature. The DMs are taken from the supplementary information Table 1. The color map is linearly scaled. and darker patches represent higher intensities. The bad frequency channels are masked and labeled using red patches on the left. The time and frequency resolutions for the plots are downsampled to 0.786 ms and 3.91 MHz, respectively. }
\end{figure*}

\clearpage

\begin{figure*}
\centering
\includegraphics[width=0.95\textwidth]{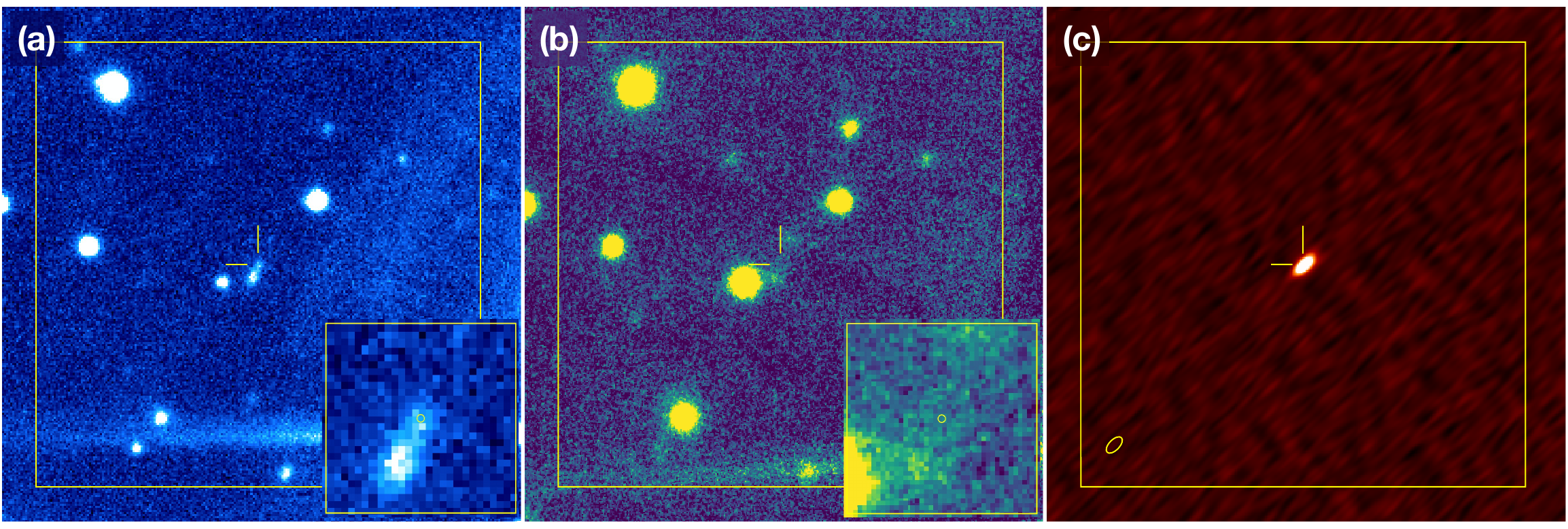}

\caption{
\footnotesize \textbf{Optical, infrared, and radio images of the field of \frb.} In each case, the box is 40\arcsec\ in size and the 2\arcsec\ crosshairs indicate the best FRB position at R.A. = 16h02m04.272s, Decl. = $-11$\arcdeg17\arcmin17.32\arcsec (J2000). North is up and East is to the left.
(a) An optical $R^{\prime}$-band image obtained by CFHT MegaCam covers 5427\AA--7041\AA, including redshifted H$\beta$\,4861\AA, [OIII]4959\AA, and [OIII]5007\AA\ emission lines from the host galaxy. The bright streak at the lower left is an artifact caused by a bright star outside the field of view. The inset shows the host galaxy in a 5\arcsec\ region centered on the FRB position, as indicated by the yellow circle (0.1\arcsec\ radius, corresponding to the 1$\sigma$ position uncertainty).
(b) The infrared $J$-band image by Subaru/MOIRCS shows emission only at the location of the peak of the optical light profile of the host galaxy. The inset is a 5\arcsec\ region matching the inset in panel (a).
(c) The radio VLA image (2--4~GHz) shows a compact persistent source at the FRB location. The synthesized beam is shown as an ellipse of size (0\farcs92 $\times$ 0\farcs47) in the left corner.
}
\end{figure*}

\clearpage

\begin{figure*}
\centering
\includegraphics[width=0.9\textwidth]{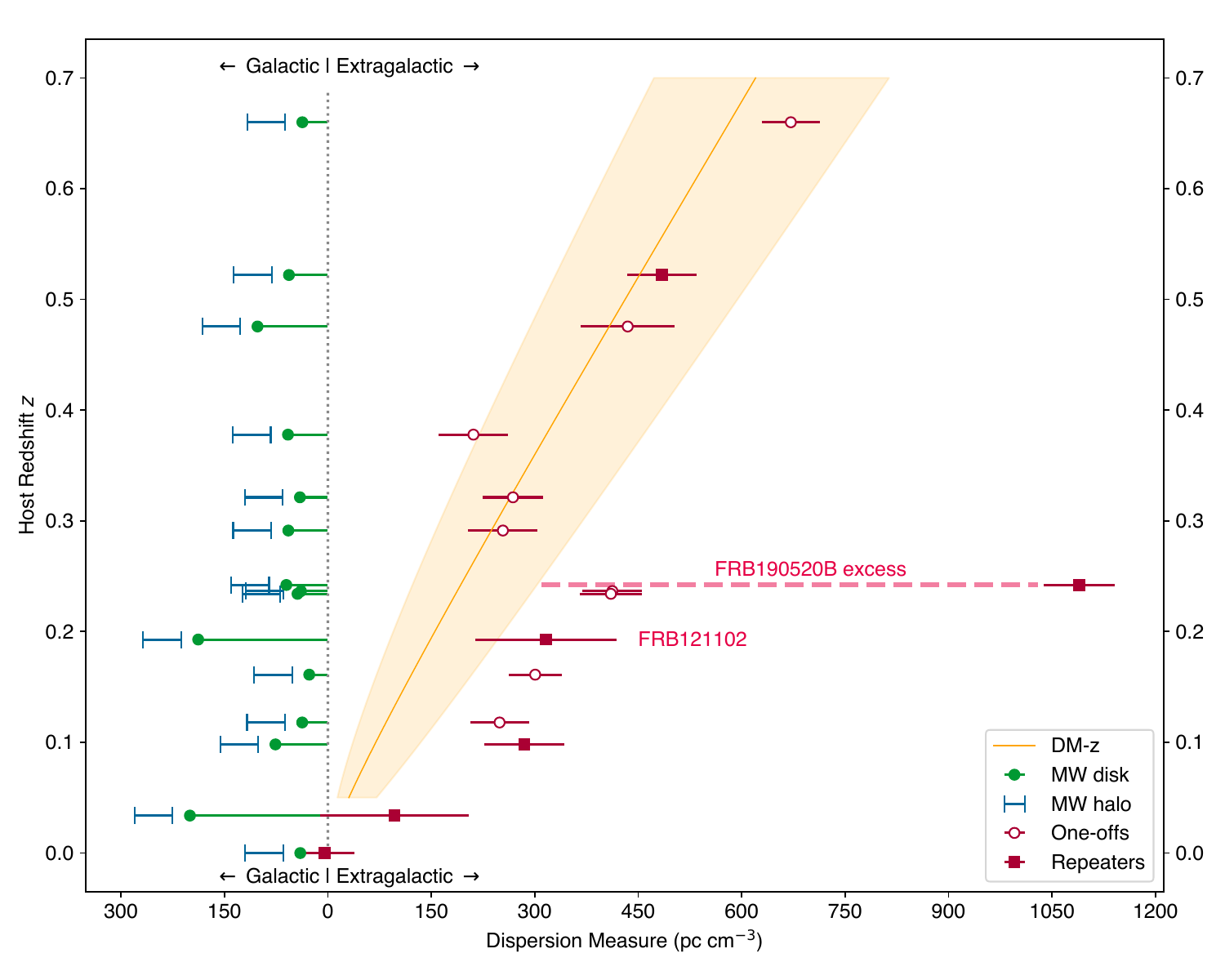}
\caption{
\footnotesize \textbf{Galactic and extragalactic contributions to the DM observed for 15 FRBs.} Theses 15 FRBs are with firm host galaxy associations and redshifts. Sources localized on initial detection (``one-off'' bursts) and repeating FRBs are identified separately.
Galactic disk contributions are estimated from NE2001\cite{Cordes03} with 
$\pm20$\%
uncertainties, along with an additional halo contribution of 25$-$80~\dmu\ (full range). Red error bars on each extragalactic DM estimate represent a conservative full range uncertainty derived by adding the disk and halo ranges.
The expected median DM contribution of the intergalactic medium and the inner 68\% confidence interval are shown by the orange line and shaded region (see Methods).
The host galaxy contributions $\rm{DM}_{\rm host}$ shift observed values to the right of the band of extragalactic DM predicted for the intergalactic medium alone. \frb\ is a clear outlier from the general trend, with an unprecedented DM contribution from its host galaxy. FRB 20121102A is identified for comparison.}
\end{figure*}

\begin{table*}
 \footnotesize
\begin{center}
\caption{Properties of FRB 20190520B}\label{tab:generaltab}
\begin{threeparttable}
\colorbox{gray!10}{%
\begin{tabular}{ll}
\hline \hline
 
 \multicolumn{2}{l}{\centering\textbf{Burst Parameters}}\\

Right Ascension (J2000)           & 16$^\mathrm{h}$02$^\mathrm{m}$04.272$^\mathrm{s}$ (0.1$^{\prime\prime}$) \\
Declination (J2000)               & $-$11$^\circ$17$^\prime$17.32$^{\prime\prime}$ $(0.08^{\prime\prime})$  \\
Galactic Coordinates ($l,b$)      & $359^\circ.67,29^\circ.91$ \\
Number of detections\tnote{a}               & 88 \\
Mean total dispersion measure (\dmu)         &
$1204.7\pm4.0$ \\
Mean burst width (ms)               & 13.5$\pm1.2$\tnote{b} \\
Mean scattering timescale~(ms)   at 1.25 GHz &  10$~\pm~$2\tnote{b}   \\
Measured fluence (Jy ms)          & 0.03 to 0.33  \\
$\rm DM_{MW,disk}, DM_{MW,halo}$ (\dmu)\tnote{c}    &  60 , 50  \\
$\rm DM_{host}$~(\dmu) & $903^{+72}_{-111}$ \\
Luminosity distance (Mpc)    &  1218  \\
Isotropic equivalent energy ($10^{37}\,\rm erg$) & 3.6 to 40\\

 \multicolumn{2}{l}{ \centering\textbf{Persistent Radio Source}}\\
Right Ascension (J2000)  & 16$^\mathrm{h}$02$^\mathrm{m}$04.261$^\mathrm{s}$ (0.1$^{\prime\prime}$)\\
Declination (J2000)  & $-11$\arcdeg17\arcmin17.35\arcsec $(0.05^{\prime\prime})$ \\
Flux density at 3.0\,GHz ($\mu$Jy) & 202$~\pm$~8\tnote{d}\\
Luminosity at 3.0\,GHz (erg s$^{-1}$ Hz$^{-1}$) &  3~$ \times~10^{29}$ \\
Size (3 GHz) & $<0.36$\arcsec\ ($<1.4$ kpc)\\
 
\multicolumn{2}{l}{\centering \textbf{Host galaxy}}\\

Redshift ($z$)         & $0.241~\pm~0.001$ \\
Stellar mass (M$_{\odot}$)\tnote{e}            & $\sim 6 \times 10^8$\\
H$\alpha$ luminosity ($\rm erg~s^{-1}$)             & $7.4\pm0.2 \times 10^{40}$\\
SFR (M$_{\odot}$ yr$^{-1}$)\tnote{f}        & $\sim0.41$ \\
\hline \hline
\end{tabular}
}
\begin{tablenotes}
\scriptsize
    \item[a] Including the FAST and VLA observations.
    \item[b] These are the mean values from FAST bursts fitted with a Gaussian pulse convolved with a one-sided exponential (see Methods).
    \item[c] The MW electron density model from NE2001 and YMW16.
    \item[d] Averaged over span of $\sim$~2 months from August 30 to November 16 of 2020.
    \item[e] Based on the $J$-band magnitude.
    \item[f] Based on the H$\alpha$ luminosity.
\end{tablenotes}

\end{threeparttable}
\end{center}

\end{table*}

\clearpage
\bibliographystyle{naturemag}


\clearpage

\noindent \textbf{Main-text Figure Legends}

\begin{enumerate}[label=\textbf{Figure \arabic*},ref=Figure \arabic*, wide]

\item \textbf{Bursts from FRB 20190520B, shown as dynamic spectra and burst profiles.} Top panel: Frequency integrated burst normalized intensities that were detected during FAST observations. These six bursts are chosen from different observation epochs. The burst labels `PXX' are in order of arrival time and the corresponding burst properties are shown in supplementary information Table 1.\label{fig:Water-Fall plots} 
Bottom panel:  De-dispersed dynamic spectra of the bursts clearly showing their band-limited nature. The DMs are taken from the supplementary information Table 1. The color map is linearly scaled. and darker patches represent higher intensities. The bad frequency channels are masked and labeled using red patches on the left. The time and frequency resolutions for the plots are downsampled to 0.786 ms and 3.91 MHz, respectively.

\item \textbf{Optical, infrared, and radio images of the field of \frb.} In each case, the box is 40\arcsec\ in size and the 2\arcsec\ crosshairs indicate the best FRB position at R.A. = 16h02m04.272s, Decl. = $-11$\arcdeg17\arcmin17.32\arcsec (J2000). North is up and East is to the left.
(a) An optical $R^{\prime}$-band image obtained by CFHT MegaCam covers 5427\AA--7041\AA, including redshifted H$\beta$\,4861\AA, [OIII]4959\AA, and [OIII]5007\AA\ emission lines from the host galaxy. The bright streak at the lower left is an artifact caused by a bright star outside the field of view. The inset shows the host galaxy in a 5\arcsec\ region centered on the FRB position, as indicated by the yellow circle (0.1\arcsec\ radius, corresponding to the 1$\sigma$ position uncertainty).
(b) The infrared $J$-band image by Subaru/MOIRCS shows emission only at the location of the peak of the optical light profile of the host galaxy. The inset is a 5\arcsec\ region matching the inset in panel (a).
(c) The radio VLA image (2--4~GHz) shows a compact persistent source at the FRB location. The synthesized beam is shown as an ellipse of size (0\farcs92 $\times$ 0\farcs47) in the left corner.\label{fig:sky}

\item \textbf{Galactic and extragalactic contributions to the DM observed for 15 FRBs.} Theses 15 FRBs are with firm host galaxy associations and redshifts. Sources localized on initial detection (``one-off'' bursts) and repeating FRBs are identified separately.
Galactic disk contributions are estimated from NE2001\cite{Cordes03} with 
$\pm20$\%
uncertainties, along with an additional halo contribution of 25$-$80~\dmu\ (full range). Red error bars on each extragalactic DM estimate represent a conservative full range uncertainty derived by adding the disk and halo ranges.
The expected median DM contribution of the intergalactic medium and the inner 68\% confidence interval are shown by the orange line and shaded region (see Methods).
The host galaxy contributions $\rm{DM}_{\rm host}$ shift observed values to the right of the band of extragalactic DM predicted for the intergalactic medium alone. \frb\ is a clear outlier from the general trend, with an unprecedented DM contribution from its host galaxy. FRB 20121102A is identified for comparison.\label{fig:scatter_dm_z}

\end{enumerate}

\pagestyle{empty}
\renewcommand{\thesubsection}{\Alph{subsection}}
\renewcommand{\thefigure}{\thesubsection .\arabic{figure}}
\renewcommand{\thetable}{\thesubsection .\arabic{table}}
\setcounter{figure}{0}
\setcounter{table}{0}
\newpage

\begin{methods}

\section*{Observations}
\subsection{FAST}

CRAFTS is a multi-purpose drift scan survey conducted with FAST using a 19-beam receiver operating at 1.05--1.45~GHz, deployed in May 2018 and conducting blind FRB searches using multiple pipelines\cite{Niu_2021}. 
\frb\ was discovered on November 16th 2019 in archived CRAFTS data which is in 1-bit filterbank format with 196 $\mu$s time resolution and 0.122 MHz frequency resolution.
In this first discovery observation, 3 bursts were detected in 10 seconds, and another burst was detected 20 seconds later. These 4 bursts from the drift scan survey gave a preliminary location for the source within a $\sim 5 $ arcmin diameter region. Taking the pointing location from FAST, 2 follow-up observations were performed with FAST on April 25th and May 22nd in 2020 with 19-beam mode in which 15 bursts were detected. A monthly observation campaign was then conducted by FAST using the $\sim 100~\rm{mas}$ localization from VLA (next section). After some regular telescope maintenance, 10 observations were performed spanning from July 30th 2020 to September 19th 2020 in which 60 more bursts were detected. 

Bursts were detected from \frb\ in each FAST monitoring observation. We list the properties of those bursts detected by FAST in supplementary information Table 1. The burst arrival time is in MJD format and has been transformed to the arrival time at the solar system barycentre (SSB) at 1.5 GHz with the DM values from supplementary information Table 1. 
The observed $\rm DM_{obs}$ is measured using the method from ref.~\cite{Hessels2019} and code from ref.~\cite{2019ascl.soft10004S}. 
We use the DM value that maximizes the structure of the highest S/N burst on a given date, and report this same DM for all other bursts detected on that date. Apparent epoch-to-epoch DM variations may simply be the result of a combination of variable effects, including intrinsic pulse structure, time$-$frequency drift, flux distribution in frequency, and scattering (which affects bursts chromatically).

The FAST  data were searched using a \texttt{Heimdall} based pipeline\cite{2012MNRAS.422..379B, Niu_2021}. For the FRB blind search in the 19-beam drift scan survey, the polarizations were added and only the total intensity (Stokes I) was recorded. The trial DM range was from 100 to 5000\dmu~and we matched the pulse width using a boxcar search up to 200 ms. The candidates above a S/N of 7 and present in less than 4 adjacent beams were manually examined for further inspection. After we identified \frb\ as a new FRB source, the follow-up burst search was done with a narrower DM range (100$-$2000\dmu). Following the VLA localization, the tracking observations only recorded data from the central beam but with all the Stokes parameters and higher time resolution ($\sim50\, \mu$s). Candidates with S/N$>$7 correspond to a fluence threshold of 9 mJy ms for a burst width of 1 ms. 

The pulse widths were estimated by a Gaussian profile fit if the burst showed no evidence of scattering (see section on scattering below). A sub pulse is recognized if the bridge between two closely spaced peaks drops more than $5\sigma$ below the higher peak. If the burst shows a scattering tail, the pulse width is derived from a combined fit for the Gaussian width and scattering time. We roughly estimate the bandwidth of each burst by dividing the whole bandpass into 50 MHz subbands and identifying the subbands containing burst emission. The burst fluences were determined using the bandwidth, temporal width, and amplitude of each burst.

\subsection{VLA}\label{method:VLA}
Following the FAST detection of bursts from \frb, VLA observations were performed from July-November 2020 (DDT project 20A-557) at the most reliable position determined using FAST (good to 5 arcmin).
Most of the observations were in the B array configuration (with a maximum baseline of 11.1~km), with the exceptions that the array configuration was BnA on MJD 59161 and BnA-$>$A on MJD 59167 and 59169. In total, 11.4~hrs were spent on the source at multiple bands. The total bandwidths at L (1.5 GHz), S (3 GHz), and C (5.5 GHz) bands were 1024, 2022, and 2022 MHz, respectively, with 1024 channels, corresponding to channel bandwidths of approximately 1, 2, and 2 MHz.

The details of the observations are given in Extended Data (ED)~Table 1.
The telescopes were pointed at the field centered at (R.A., Decl.)[J2000] = (16h02m01s, $-11$d17m28s). 
We used the \realf\ search system at VLA to search for bursts from \frb\ in our VLA observations. Observations on MJD 59169 contributed to the imaging at S band, but the \realf\ system was not run on this day due to a system error. 

The \realf\ search system has been described in detail in refs\cite{law18, law2020}, but here we discuss it briefly. Using a commensal correlator mode, the visibilites are sampled with 10~ms resolution and distributed to the \realf\ GPU cluster to search for transients. The search pipeline \texttt{rfpipe}\cite{rfpipe} then applies the online calibration, de-disperses the visibilities, and forms images at a set of different temporal widths. 
The 8$\sigma$ fluence limit of a 10 ms image is 0.29, 0.18 and 0.13~Jy\,ms at L, S and C band, respectively. Candidates with image S/N greater than the threshold triggered the recording of fast sampled visibilities in windows ranging from 2 to 5 seconds centered on each candidate. For each candidate, time-frequency data corresponding to the maximum pixel in the image are then extracted and post-processed. The candidate is then classified by \texttt{fetch}, a GPU based convolutional neural network\cite{agarwal2020}.

Promising candidates selected visually by the \realf\ team go through an offline analysis to refine the candidate parameters and improve its detection significance, if possible. Several methods are tried within this analysis: offline search for the transient using a finer DM grid, varying RFI flagging thresholds, changing image size (as the realtime-search is done on non-optimal image sizes due to computational limitations), sub-band search etc. (see Section 2.4 of ref.~\cite{law2020} for more details). The pixel sizes of images formed by the realtime system at L, S, and C band were 0.9\arcsec, 0.48\arcsec, and 0.27\arcsec, respectively, with an image size of 2048$\times$2048 pixels corresponding to field sizes of 0.5\arcdeg, 0.27\arcdeg, and 0.15\arcdeg. During refinement, we searched the data with smaller pixel sizes: 0.5\arcsec$\times$0.75\arcsec\ at L band, 0.38\arcsec$\times$0.28\arcsec\ at S band, and 0.27\arcsec$\times$0.18\arcsec\ at C band. The significance of an astrophysical transient improves when the data is de-dispersed at a DM closer to the true DM of the candidate (see Section 2 of ref.~\cite{rfclustering}). Therefore, we re-ran the search with a finer DM grid at 0.1\% fractional sensitivity loss, as compared to 5\% used by the realtime system. Noise-like events or RFI are sensitive to RFI flagging and image gridding parameters, and so they cannot be reproduced on refinement and are discarded. Sometimes the transient signal is present only in a fraction of the whole frequency band, so we also re-run our search only using the relevant frequencies, which further improves the detection significance. We applied these techniques on all the candidates selected from these observations to refine their parameters.

\section*{Localization of bursts}
\label{sec:burst_locs}

\subsection{Calibration}

The \realf\ system makes several assumptions during calibration and imaging in order to improve computational efficiency for the real-time imaging and search.
Moreover, the Point Spread Function (PSF) of the interferometer is not deconvolved from the image, so the real-time system forms ``dirty" images. The PSF shape makes the images more difficult to visually interpret and model.

To address these issues, we used a top-hat function to select the raw, de-dispersed, fast-sampled visibilities containing only the burst signal to re-image the burst data using the Common Astronomy Software Application package (CASA v5.6.2-3)\cite{2007ASPC..376..127M}. 
Observations of 3C\,286 (acquired before the FRB observations) were used to calibrate the flux density scale, bandpass, and delays. Complex gain fluctuations over time were calibrated with observations of calibrator J1558$-$1409. We performed phase-reference switching on intervals of 16, 13, and 12 minutes for L, S, and C-bands, respectively, consistent with the nominal phase-coherence timescales for the VLA, so that after calibration any short-timescale phase variations are negligible. Therefore, the systematic errors for the burst positions on short timescales are of the same magnitude as those for the deep imaging, as discussed below.

\subsection{Determining properties of individual bursts}
After calibration, the CASA task \texttt{tclean} was used to generate an image for each burst and estimate the S/N. Most of the bursts are spectrally confined, so for each burst we select the spectral window range that produces the highest image S/N.
We then use the CASA task \texttt{imfit} around the FRB position to fit an elliptical Gaussian to the source in the radio image and measure the centroid location, peak flux density and 1$\sigma$ image-plane uncertainties. Hereafter, we refer to these image-plane uncertainties as statistical errors on burst positions.

The average burst positions are calculated for each frequency band separately, by weighting each detection by the inverse of the position fit errors (statistical errors) reported by CASA. 
Statistical errors are inversely proportional to S/N, and therefore high significance detection is expected to have smaller fit errors. 
The total positional error at each frequency band is obtained by adding this statistical error in quadrature with the systematic error in that band (determined using the deep radio images described below).
The total positional error (R.A. error, Decl. error) is (0.25$^{\prime\prime}$, 0.32$^{\prime\prime}$) at L band, (0.28$^{\prime\prime}$, 0.17$^{\prime\prime}$) at S band, and (0.12$^{\prime\prime}$, 0.09$^{\prime\prime}$) at C band. In the final analysis, the errors are dominated by systematic uncertainties at each frequency band (see \ref{tab:vla_prs_locs} and \ref{tab:vla_burst_locs}).
The final burst position was obtained by taking a weighted average of the burst positions at each of the three frequency bands, using the inverse of the total error at each frequency band as the weight.
The best estimate of the burst position is R.A.= 16h02m04.272s, Decl.= $-11$\arcdeg17\arcmin17.32\arcsec (J2000). We estimate the error on this position to be 0.10$^{\prime\prime}$ and 0.08$^{\prime\prime}$. We calculated the reduced chi-square value of individual burst positions with respect to this best estimate of the FRB position. The reduced chi-square was given by $\chi^{2}=(1/8)\sum_{i} \left(\theta_{i}-\theta_\mathrm{best}\right)^{2}/{\sigma_{i}^{2}}$, where $\theta_i$ refers to the coordinates of a burst, $\theta_\mathrm{best}$ refers to the best estimate of FRB position, $\sigma_i$ is the total error on a burst (statistical and systematic errors added in quadrature) and the sum is over all the 9 localized bursts. We obtained a reduced chi-square value of 0.53 and 0.77 for R.A. and Decl. respectively.

We confirmed that there were no short-timescale phase errors by performing intermediate-timescale imaging of the continuum data. We imaged segments of 5-30 seconds, such that at least 1-3 sources could be detected at each frequency, and inspected the position variations of those sources over time. At all frequencies, the positions were stable and consistent with the statistical uncertainty (radiometer noise). Offsets of the sources were within the range of the systematics quoted for the PRS in next section. 

Due to the low time resolution of VLA data (10 ms), we do not perform any sophisticated modeling of burst properties. The properties of the VLA bursts, including the flux densities, were obtained using CASA and the \realf\ system, and are given in \ref{tab:vla_burst_props}. Here, the reported widths should be considered as upper limits. The frequency extent of the burst signal was visually determined and reported in the last column. We also could not estimate a structure maximizing DM for VLA bursts because of the coarse time resolution. The DMs reported in this table are the values that maximize the S/N. This also explains the apparent variability seen in the DM values. 

\section*{Persistent Radio Source}

\subsection{Deep radio images and PRS}\label{sec:prs}
The VLA campaign obtained two epochs at 1.5 GHz and six epochs at 3 GHz and 5.5 GHz, which resulted in an on-source time of $\sim$ 3 hours for 1.5 GHz, and $\sim$ 4 hours for both 3 GHz and 5.5 GHz.

Along with the \realf\ output, the VLA visibilities with 3~s or 5~s sampling times 
were saved and analyzed to search for persistent radio emission. This is done in parallel to saving the high time resolution (10~ms) data around the burst that was used in the previous section. We used the same data reduction, flagging, and calibration approach that was used for burst localization.
We then performed further flagging on the target and then subsequently imaged its Stokes I data using the CASA deconvolution algorithm \texttt{tclean}. To balance sensitivity while reducing sidelobes from a nearby bright source, we imaged with a Briggs weighing scheme (robust=0). In addition, self-calibrations were performed for all observations to correct considerable artefacts from the close-by bright sources in the field. We made use of the CASA task \texttt{imfit} to measure source flux densities by fitting an elliptical Gaussian model in the image plane.

We stacked observations at each central frequency in the uv-plane and then imaged the Stokes I intensity, resulting in deep images at 1.5 GHz, 3 GHz, and 5.5 GHz with rms noise of 9.0, 4.5, and 3.0 $\mu$Jy/beam, respectively. At 1.5 GHz, 3 GHz, and 5.5 GHz, the upper limits of the deconvolved sizes of the PRS are as large as 1.4\arcsec$\times$0.89\arcsec, 0.51\arcsec$\times$0.14\arcsec, and 0.36\arcsec$\times$0.1\arcsec, respectively. 
Thus a conservative upper limit of the size of the PRS from VLA observations is 0.36\arcsec, which corresponds to 1.4 kpc at the angular diameter distance of 809 Mpc. The obtained positions for the PRS are shown in \ref{tab:vla_prs_locs} and the positions of the bursts and PRS are shown in \ref{fig:astrometry}.
The systematic offsets on these positions are estimated in the next section. 

\subsection{Systematic Offsets} 

In order to determine the systematic errors on the coordinates of the PRS that we determined from the deep images, we ran the \texttt{PyBDSF} (\url{https://www.astron.nl/citt/pybdsf/index.html}) package to extract radio sources from the deep images. We then cross-matched the detected point sources in the deep images with the sources listed in the optical PanSTARRS survey DR1\cite{panstarrs}. We identified the radio point sources using the following criteria:
\begin{itemize}
 \item The peak intensity (Jy/beam) of a source should be 0.7, 0.5, 0.5 times higher than its integrated flux (Jy) for the 1.5 GHz, 3 GHz and 5.5 GHz images, respectively.
 \item The S/N (peak intensity / local rms noise) of a source should be greater than 5. 
\end{itemize}

In total, we detected 375, 113, and 43 sources in the 1.5 GHz, 3 GHz, and 5.5 GHz deep image, respectively. The selected sources were also visually checked to make sure that they are `point-like' sources in the deep images. We searched for matching optical sources within radii of 0.5\arcsec, 1.0\arcsec, and 2.0\arcsec. 
Going from 1.0\arcsec\ to 2.0\arcsec, the additional cross-matched sources at each band are consistent with chance coincidence. Therefore, we adopt a 1.0\arcsec\ cross-match radius, finding 136, 31, 9 sources with optical counterparts in the PanSTARRS DR1 catalogue at 1.5 GHz, 3 GHz, and 5.5 GHz, respectively. These cross-matched sources were used to determine the astrometry for FRB 20190520B and its PRS. By subtracting PanSTARRS coordinates from VLA coordinates, we estimate the systematic offsets in the 1.5 GHz, 3 GHz and 5.5 GHz positions, as listed in \ref{tab:vla_prs_locs}. 
The systematic offsets are consistent with zero mean, and their uncertainties dominate the uncertainties of the PRS position. 

\subsection{Determining the position of the PRS}
To determine the positions of the PRS in the three frequency bands, we followed a procedure similar to what was used to determine the burst positions (see Section~\nameref{sec:burst_locs}). 
The best estimate for the PRS position is (R.A., Decl.) [J2000] = (16h02m04.261s, $-11$\arcdeg17\arcmin17.35\arcsec). The error on this position is estimated to be 0.10$^{\prime\prime}$ and 0.05$^{\prime\prime}$ or R.A. and Decl., respectively.

\subsection{Variability and Spectrum of the PRS} 
The flux density of the source measured at each epoch is shown in \ref{tab:vla_observations}.
Measured flux densities show variations that are mostly consistent with measurement errors, but there are $\sim2\sigma$ variations in the 2.5 GHz subband that, if real, could be refractive interstellar scintillation, or intrinsic variations, or both.
In order to study the spectrum of the PRS, we split each of the observations into two, 0.5~GHz/1~GHz sub-bands. Then we measured the average flux density at each of the subbands over the campaign. 
The multi-band data were fit with a power-law model ($S_{\nu} \propto \nu^{\alpha}$, where $S_{\nu}$ is the observed flux density at the frequency $\nu$, $\alpha$ is the spectral index), yielding an average spectral index for the PRS of $-0.41\pm$0.04 (see \ref{fig:spectrum}). 

\subsection{Chance coincidence association of the PRS}
In the 1.5 GHz deep image, we detected 8 `point-like' sources, including the PRS,  within 5 arcminutes of the phase center and with a flux density no less than that of the PRS (at 260 $\mu {\rm Jy}$) based on our point-source selection criteria described earlier. There is an additional bright source with a flux density of a few tens of mJy in the region, but it was not classified as a `point-like' source based on our criteria.

To estimate the chance coincidence probability of the PRS with the bursts, we compared the solid angle corresponding to the uncertainty of burst localization and the average solid angle occupied by each of the eight `point-like' sources in the FRB field-of-view. The solid angle corresponding to each of the 8 `point-like' sources is roughly estimated as S$_{\mathrm{source}}$= $\pi (5/60)^{2}/8$ Sr. The offset between the average position of the nine bursts and the position of the PRS at 5.5 GHz, which is best constrained when taking both statistical and systematic errors into account, is about 0.06 arcseconds. This, along with a statistical error of 0.01 arcseconds and a conservative estimate of the systematic error of 0.12 arcseconds, can be used to estimate the offset between the PRS and the FRB position. We conservatively estimate this offset to be 0.19 arcseconds. The solid angle corresponding to the offset therefore can be estimated as S$_{\mathrm{offset}}$=$\pi \times (0.19/60/60)^{2}$ Sr. The ratio between S$_{\mathrm{offset}}$ and S$_{\mathrm{source}}$ gives the chance of coincident association of the PRS with the FRB position to be $\approx3\times{10}^{-6}$.

\section*{Galaxy Photometry and Redshift Determination}
The deep $R^{\prime}$-band (5427\AA - 7041\AA) 
images obtained by CFHT/MegaCam were stacked from archival observation data taken in 2014-2015 by the CFHT archival pipeline MEGAPIPE, with a total of $\sim$ 3.6 hours on the field.
The level 3 (flux calibrated) images were retrieved for our analysis.

NIR $J$-band images of the \frb\ field were taken under a relatively poor seeing condition ($\sim 1.3\arcsec$) through a Subaru target-of-opportunity observation on August 5th 2020. A total of 1.4-hour observations were used for the final combined $J$-band image shown in \ref{fig:sky}. A $J$-band source of 22.07$\pm$0.14 mag (AB) was detected at $\sim 1\arcsec$ south east of the burst location, and is possibly stellar emission from the host galaxy. A faint source 2.5\arcsec\ north has 22.87$\pm$0.26 mag in $J$-band. Neither of these two sources was detected in the $Ks$-band image, with a $5\sigma$ limit of 21.74 mag (1.1 hours). 

An optical spectrum was obtained with the Double Spectrograph (DBSP) on the Palomar 200-inch telescope on 24th July 2020 using a 1\arcsec\ slit-width. This observation was executed before the CFHT/MegaCAM archival data on \frb\ field were found, and only Pan-STARRS images were used for observation planning. 
The slit of DBSP was set to cover the PRS emission at R.A. = 16h02m04.27s; Decl. = $-11^\circ17^\prime17.5^{\prime\prime}$ detected by VLA in L-band on 22nd July 2020. 
No clear optical counterpart was detected in any of the 5 band images of PanSTARRS from DR1 \cite{panstarrs}. The slit was guided by the nearby M-star at R.A. = 16h02m04.48s, Decl. = $-11^\circ17^\prime19.1^{\prime\prime}$ as reported in the PanSTARRS DR1 catalog, with $i$ = 20.4 mag, and 3.4\arcsec\ due east of the PRS coordinates.
The slit was set to a position angle of 108.5 degrees, ensuring that both the PRS and the M-star fell within the slit. The observations with 2$\times$900 s exposures were carried out under photometric sky conditions and sub-arcsecond seeing. The 2D spectrum was generated using IRAF, including bias removal, flat-fielding, and reduction of other instrumental effects. The 1D spectrum was extracted from a 1.5\arcsec\ window. The standard star BD$+$28 4211 was used for telluric correction and flux calibration. 
The DBSP 1D spectrum is shown in \ref{fig:spectrum_line}. The flux scale of the spectrum does not include the slit loss and registration error of Pan-STARRS coordinates of the M-type star. The [OIII] 5007\AA\ line and the $\Ha$ line are both well detected ($> 5 \sigma$). The two emission lines are narrow, each with a FWHM of $\sim$ 10\AA. The redshift derived from these two spectral lines is $z$ = 0.241$\pm0.001$. 

A follow-up Keck LRIS spectroscopic observation was carried out on 25th August 2020 under reasonable weather and seeing conditions (1.1\arcsec) . The 1.5\arcsec\ slit was set at a position angle of $160^{\circ}$ to the extended optical emission seen in the MegaCam $R^{\prime}$-band image around the \frb\ location. A total exposure of 3600\,s was obtained. 
The emission lines $\Ha$, H$\beta$, [OIII]4859\AA, and [OIII]5007\AA~are well detected, indicating that the extended $R^{\prime}$-band structure has the same redshift of $z = 0.241$.
After the spectrum was corrected for Galactic foreground extinction, the line flux ratio between $\Ha$ and H$\beta$ was used to estimate an extinction of $A_{V} = 0.80$ mag, assuming a case-B line ratio of 2.86, yielding an H$\alpha$ flux $F_{\rm H\alpha} = (42.0 \pm 0.5) \times 10^{-17}$ erg s$^{-1}$ cm$^{-2}$. The extinction-corrected $\Ha$ luminosity is $L_{H{\alpha}} = 7.4\pm0.2 \times 10^{40}$~erg s$^{-1}$.

\subsection{Chance association probability of the host galaxy}
We use the approach of ref.~\cite{2002AJ....123.1111B} to estimate the chance coincidence probability of the association between the galaxy (J160204.31$-$111718.5) and \frb. Assuming a uniform surface distribution of galaxies, the probability of chance coincidence follows a Poisson distribution; i.e., $P = 1 - \exp(-n_i)$, where $n_i$ is the mean number density of galaxies brighter than a specified $R^{\prime}$-band magnitude in a circle of given radius, determined by the half light radius of the galaxy and the burst localization error region. This number density is estimated using the results from ref.~\cite{hogg1997}. 
From previous discussion, we assume the localization error on the FRB position to be $\sim$ $0\farcs1$, and the size of the host galaxy to be $0\farcs5$.
The $R^{\prime}$-band magnitude of the possible host is difficult to estimate since it is significantly affected by the emission lines. Thus we conservatively estimate the $R^{\prime}$-band magnitude to be $\lesssim$ 23.3 mag assuming a flat spectral energy distribution in $\nu L_{\nu}$ between $R^{\prime}$-band and $J$-band, where $\nu$ and $L_{\nu}$ are frequency and luminosity, respectively. The probability of a chance association between \frb\ and J160204.31$-$111718.5 is estimated to be $\sim$ 0.8$\%$, independent of constraints from the association with a rare PRS or the observed FRB scattering. The next nearest galaxy is 6.5\arcsec\ away from the FRB and has a chance coincidence probability of $>20\%$.

\section*{FAST Burst Sample Analysis}

\subsection{Repetition rate}

\frb\ was found to be 
regularly active with two or more bursts  detected in each monitoring observation with FAST. Because of the possible clustering behavior of FRB emission, a Weibull distribution was employed
for the waiting time $\delta$ (the separation between adjacent detected bursts) \cite{2018MNRAS.475.5109O}, 
\begin{equation}
 \mathcal{W} = \frac{k}{\lambda}\left(\frac{\delta}{\lambda}\right)^{k-1}e^{-(\delta/\lambda)^k},
\end{equation}
where $k$ is the shape parameter and  $\lambda$ is the scale parameter of the distribution.
The Weibull distribution reduces to a Poisson distribution when $k=1$.
The mean waiting time, $E(\delta)=\lambda\Gamma(1+1/k)$,
implies a burst rate  $r=1/E(\delta)$. \ref{fig:weibull} shows the inferred parameter distributions obtained with a Markov Chain Monte Carlo (MCMC). We find the burst rate of \frb{} is $r=4.5^{+1.9}_{-1.5}\ \rm hr^{-1}$ with shape parameters $k=0.37^{+0.04}_{-0.04}$ when all 79 bursts above 9 mJy ms are used (left panel in \ref{fig:weibull}). Waiting times shorter than 1~s include substructure in individual bursts, so we also analyze only longer wait times $\delta \ge 1$~s, yielding    $r=5.3^{+1.1}_{-1.0}\ \rm hr^{-1}$ with shape parameters $k=0.76^{+0.09}_{-0.08}$
(right panel in \ref{fig:weibull}).

\subsection{Short and long timescale periodicity search}

For a range of trial periods ($P$) and period derivatives ($\dot P$), the times-of-arrival (ToAs) of the 79 FAST bursts were folded to calculate the phase of each burst for a given trial period. For the short timescale periodicity search, the ToAs were folded at $P$ between 1~ms and 1000~s and $\dot P$ between $10^{-12}$ and $1$ s s$^{-1}$. For the long timescale periodicity search, the trial range of period $P$ was from 2 to 365~d and the range for period derivative (to account for spindown) was $\dot P$ from $10^{-12}$ to 1~d d$^{-1}$. The longest contiguous phase segment without bursts was calculated for each fold trial as a signature for possible periodicity.

For the short timescale periodicity search, the bursts spread to a phase window larger than 60\% of one period, which indicates no periodicity pattern in the trial range. For the long timescale periodicity search, the longest phase segment without bursts showed two maxima $\sim0.6$ near 67~d and 169~d. These two maxima are reproduced by folding the MJDs of all observation sessions, which indicates that the maxima reflect the sampling window rather than any burst periodicity. Thus, no long or short period of~\frb~was detected.

\subsection{Energy distribution}

A 1-Kelvin equivalent noise calibration signal was injected before each observation session to obtain a high quality flux density and energy calibration measurement for each detected burst. The injected noise calibration signal was used to scale the data into temperature units, yielding a nearly constant rms radiometer noise to within $6\%$.
We then converted from units of Kelvin to Jy using the zenith angle-dependent gain curve, provided by the observatory through quasar measurements \cite{Jiang_2020}. For most observations, the zenith angle is $<$ 20 degrees, which corresponds to a stable gain of $\sim$16 K/Jy.

Assuming bursts have spectral index of about 0, we
calculated their isotropic equivalent burst energies, $E$, following Equation (9) of ref.~\cite{2018ApJ...867L..21Z}:
\begin{equation}
E = 10^{39}  {\rm erg} \times \frac{4\pi}{1+z} \left(\frac{D_L}{10^{28}{\rm cm}}\right)^{2}
\left(\frac{F_{\nu}}{\rm Jy\cdot ms}\right)
\left(\frac{\nu_c}{\rm GHz}\right),
\end{equation}
where $F_{\nu}$ = $S_{\nu}\times W_{\rm eq}$ is the specific fluence in units of ${\rm erg \ cm^{-2} Hz^{-1}}$ or $\rm Jy \cdot ms$, 
$S_{\nu}$ is the peak flux density which has been calibrated with the noise level of the baseline, giving the flux measurement for each pulse at a central frequency of $\nu_c$ = 1.25 GHz, $W_{\rm eq}$ is the equivalent burst duration, and the luminosity distance $D_L$ = 1218 Mpc corresponds to a redshift $z$ = 0.241 for the source of \frb, using standard cosmological parameters \cite{Planck20}. The fluence-width distribution at 1.25 GHz for \frb~bursts can be seen in \ref{fig:fluence}. The histogram of burst energies (\ref{fig:enhist}) exhibits a bump that we fit with a log-normal function.

\subsection{DM inventory analysis}

The observed DM can be separated into four primary components (all in the observer's frame):
\begin{equation}
\rm DM_{obs}=DM_{MW}+\DMmwh+DM_{IGM}+DM_{host}
\end{equation}
where $\rm DM_{MW}$ is the contribution from the Milky Way's interstellar medium, $\rm DM_{halo}$ is the contribution from the Milky Way halo, $\rm DM_{host}$ the contribution from the host galaxy including its halo and any gas local to the FRB source, and $\rm DM_{IGM}$ is the contribution from the intergalactic medium. 

We use the NE2001 model\cite{Cordes03} to evaluate $\DMm = 60~\DMunits$ (compared to 50~$\DMunits$ from the YMW16 model\cite{Yao17}) as the mean of a uniform distribution with a generous $\pm40$\% width to conservatively represent the uncertainty in $\DMm$ for the direction of  \frb. For the MW halo contribution, we use a uniform distribution from 25 to 80~$\DMunits$. Together the MW disk and halo components yield a total
range from 61 to 164~$\DMunits$ with a mean of 113~$\DMunits$ and RMS uncertainty of 17~$\DMunits$.

For the IGM we use the $\Lambda$CDM cosmological model to calculate the mean DM contribution\cite{Deng14},
\begin{equation}
{\rm DM}_{\rm IGM}(z) = \frac{3cH_0\Omega_b f_{\rm IGM}}{8\pi G m_p}\int^z_0\frac{\chi(z)(1+z)dz}{[\Omega_m(1+z)^3+\Omega_{\Lambda}]^{\frac{1}{2}}},
\label{eq:IGM}
\end{equation}
where the free electron number per baryon in the universe is $\chi(z) \approx 7/8$, (assumed constant for FRB redshifts), 
the normalized matter density is
$\Omega_m =0.315 \pm 0.007$, the dark energy fraction is 
$\Omega_{\Lambda} \simeq 1 - \Omega_m$, the baryonic fraction is $\Omega_b = h^{-2} \times(0.02237~\pm~0.00015)$, the fraction of baryons in the IGM is $f_{\rm IGM}$,  and the Hubble constant is $H_0 \equiv h \times 100 \,~\rm km\,s^{-1} = 67.36\rm~\pm~0.54~\,\rm km\,s^{-1}Mpc^{-1}$ \cite{Planck20}. Using values for the speed of light $c$, the gravitational constant $G$, and the proton mass $m_{\rm p}$, the resulting expression, 
\begin{equation}
 {\rm DM}_{\rm IGM}(z) \approx 
 978~\DMunits \ f_{\rm IGM}
 \int^z_0\frac{(1+z)dz}{[\Omega_m(1+z)^3+\Omega_{\Lambda}]^{\frac{1}{2}}},
 \label{eq:dmigm}
\end{equation}
yields $\DMi(0.241) = 248 f_{\rm IGM} \, \DMunits$.

Using $\DMi(z)$ as the mean, we calculate the range of values for a given redshift  using a log-normal distribution with variance $\sigma^2_{\DMi}(z) = \DMi(z) \DM_{\rm c}$,
where $\DM_{\rm c} = 50~\DMunits$ is chosen to provide cosmic variance consistent with published simulations. This gives 
$\sigma_{\DMi}(0.241) = 111 \sqrt{f_{\rm IGM}}~\DMunits$. 
The parameters for the log-normal distribution are then
$\sigma_{\ln \DMi} = \{\ln [1 + (\sigma_{\DMi}(z) / \DMi(z))^2]\}^{1/2}$
and $\mu_{\ln \DMi} = \ln [\DMi(z)] - \sigma_{\ln \DMi}^2/2$. Note that the log-normal distribution is asymmetric.
Equation~\ref{eq:dmigm} and the log-normal distribution are also used to estimate the median $\DMi$ and its inner 68\% uncertainty range shown in \ref{fig:scatter_dm_z}.

To constrain the host-galaxy DM, we combine the MW and IGM estimates and their uncertainties with the measured DM averaged over all bursts, which is
$\rm DM_{obs}=1204.7~\pm~4.0~\DMunits$.
The cumulative distribution function (CDF) of the log-normal distribution for the IGM contribution  yields the range   $\DMi = 195^{+110}_{-70}~\DMunits$. We then
 calculate the posterior distribution of $\DMh$ after marginalizing over the IGM and Milky Way distributions and using a flat, uninformative prior for $\DMh$. The median and probable ranges are calculated from the corresponding CDF.
Using $f_{\rm IGM} = 0.85$ for the baryon fraction in the IGM\cite{1998ApJ...503..518F, 2018ApJ...867L..21Z}, we obtain a median value for $\DMh$ and 68\% probable interval
$\DMh = 903^{+72}_{-111}~\DMunits$.
When we vary $f_{\rm IGM}$ from 0.6 to 1, the total range of values for 
$\DMi$ is $\sim 80$ to 350~$\DMunits$ and for 
$\DMh$ is $\sim 1020$ to 745~$\DMunits$, where we have taken the median values resulting from each value of $f_{\rm IGM}$ and extended the ranges using (half of the) corresponding 68\% probable ranges for the smallest and largest values of $f_{\rm IGM}$.

For comparison we also give  redshift estimates for different values of $f_{\rm IGM}$ if a fixed value of $\DMh = 50~\DMunits$ is used (as is often found in the literature) along with the above quoted mean value for the MW contribution. For $f_{\rm IGM}$ ranging from 0.6 to 1, we obtain a range of redshift values from 2.2 to 0.9 (including the 68\% redshift interval derived from the redshift CDF for each value of $f_{\rm IGM}$), much larger than for the redshift of 0.241 for the  identified host galaxy.

\subsection{Constraints on $\rm{\bf{DM_{host}}}$ from H$\alpha$ Measurements}

\indent The DM of the host galaxy is independently estimated from its H$\alpha$ emission by converting the extinction-corrected H$\alpha$ flux, $F_{\rm H\alpha} = (42.0 \pm 0.5) \times 10^{-17}$ erg s$^{-1}$ cm$^{-2}$, to an H$\alpha$ surface density of  $224\pm3$ Rayleighs in the source frame at $z = 0.241$, assuming host galaxy dimensions of 0.5 by 0.5 arcseconds\cite{Ten17}. These host galaxy dimensions are only a rough estimate based on the size of the H$\alpha$ emitting region in the Keck image, but because the image is seeing limited the assumed dimensions may be even smaller, which would only serve to increase the H$\alpha$ surface density in the following calculations. The H$\alpha$ flux and estimated surface density are very similar to those found for the host galaxy of FRB 20121102A \cite{Ten17}. The H$\alpha$ surface density in the source frame $S({\rm H\alpha})_{\rm s}$ is related to the H$\alpha$ surface density in the observer frame as $S({\rm H\alpha})_{\rm s} = (1+z)^4 S({\rm H\alpha})_{\rm o}$, and is related to the emission measure (EM) in the source frame by

\begin{equation}
{\rm EM_s} = 
2.75 \ {\rm pc~cm^{-6}} \
T_4^{0.9} S({\rm H\alpha})_{\rm s}
\approx 
616 \pm 7 \ {\rm pc~cm^{-6}}
\times
T_4^{0.9} 
 \left[\frac{S({\rm H}\alpha)_{\rm s}}{224 \pm 3 \ {\rm R}}\right] ,
\end{equation}

where we have used the redshift $z = 0.241$ and $T_4$ is the temperature in units of $10^4$ K. The EM is related to the DM in the ionized cloudlet model \cite{Ten17} by ${\rm EM} = [\zeta(1+\epsilon^2) / f] {\rm DM}^2 / L$, where $\zeta$ represents cloud-cloud variations in the mean density, $\epsilon^2$ represents the variance of density fluctuations in a cloud, $f$ is the filling factor, and $L$ is the path length through the gas sampled by the FRB. Using this relation, we obtain the corresponding source frame DM,

\begin{equation}
 {\rm DM_s} 
 \approx 392\pm 3 {\ {\rm pc \ cm^{-3}} \times 
 T_4^{0.45}
 \left(\frac{L}{5 \ \rm kpc}\right)^{1/2}
 \left[\frac{f/0.1}{\zeta(1+\epsilon^2)/2}\right]^{1/2}
 \left[\frac{S({\rm H}\alpha)_{\rm s}}{224 \pm 3 \ {\rm R}}\right]^{1/2}} ,
\label{eq:DMs}
\end{equation}
where we have adopted fiducial values of $f = 0.1$ and $\zeta = \epsilon^2 = 1$, which are typical for the warm ionized medium in the Milky Way. For the path length through the gas we adopt a fiducial value of $L = 5$ kpc, which is based on the apparent size of the H$\alpha$ emitting region in the host galaxy. However, given that the optical image is seeing limited, and that the orientation of the galaxy relative to the FRB line-of-sight is not known, this path length could be as large as 10 kpc or as small as 0.1 kpc. For a range of $L$ from 0.1 to 10 kpc, we find that ${\rm DM_{s}}$ could be as small as 55 pc cm$^{-3}$ or as large as 560 pc cm$^{-3}$, for the same fiducial values of $T_4$, $f$, $\zeta$, and $\epsilon^2$. The estimated DM is also highly sensitive to the temperature: for a range of $T_4$ from 0.5 to 5, ${\rm DM_{s}}$ could range from 290 to 810 pc cm$^{-3}$, keeping all other parameters fixed at their fiducial values. In the observer's frame, the measured DM contribution of the host galaxy is smaller by a factor of $1/(1+z)$, yielding a nominal value of 
${\rm DM_{{host},coeff}} \sim 316\pm 2$ \dmu~for the coefficient in Equation~\ref{eq:DMs}. The quoted errors only account for measurement errors in the H$\alpha$ flux. To match the inferred value of ${\rm DM_{host}} \sim 900$ \dmu~requires that the three factors in Equation~\ref{eq:DMs} involving $T_4, L$ and $f$ combine to a factor $\simeq 3$, which may easily be explained by a higher temperature or the broad range of possible values for $L$. Regardless, the large H$\alpha$ EM affirms that the FRB DM receives a significant contribution from ionized gas in the host galaxy, but it is unclear whether the diffuse, H$\alpha$ emitting gas can account for the entire host galaxy contribution or whether the FRB's local environment contains significant ionized gas content that is not seen in H$\alpha$.

\subsection{Extragalactic scattering}\label{sec:scatter}

Scatter broadening manifests as a frequency-dependent temporal asymmetry in the burst dynamic spectrum, which is typically modeled as a one-sided exponential pulse broadening function convolved with a Gaussian pulse. Some bursts from \frb\ have burst structure that is suggestive of scattering: leading edges that are aligned across the radio frequency band coupled with temporal asymmetries that broaden at lower observing frequencies. However, a number of bursts are symmetric across the radio frequency band and vary significantly in burst width. When the burst S/N is low, it is also difficult to distinguish scattering from other burst substructure, such as the time-frequency drift of intensity islands\cite{Hessels2019} (often referred to as the ``sad trombone" effect). Examples of bursts with and without evidence of pulse broadening are shown in \ref{fig:scattering_examples}.

To measure the mean burst width and scattering timescale of \frb, we first integrate the dynamic spectrum of each burst along the frequency axis and normalize the resulting burst profile. Each burst profile is fit with a model comprised of a Gaussian component convolved with a one-sided exponential. Out of the 79 fitted scattering times, all scattering times with fractional uncertainties $>50\%$ are excised, leaving a subset of 26 bursts. The scattering times of these 26 bursts are then compared to their dynamic spectra and 1D burst profiles to verify that the burst temporal asymmetry is broadly consistent with a $\nu^{-4}$ frequency scaling. Table 1 in the Supplementary Information shows the scattering times re-scaled to 1.25 GHz assuming a $\nu^{-4}$ frequency dependence. The remaining 53 bursts, whose scattering times have fractional uncertainties $>50\%$, are then re-fit with a Gaussian-only model for the 1D burst profile. The full-width-at-half-maximum burst widths are also shown in supplementary information Table 1. This provisional approach yields a mean scattering time and burst width of 10 $\pm$ 2.0~ms and 13.5 $\pm$ 1.2~ms at 1.25 GHz, respectively. A more detailed study investigating the burst widths in separate frequency subbands is in progress, and corroborates the scattering interpretation presented here.

The observed mean scattering time of 10~$\pm$~2~ms at 1.25 GHz is likely too small for the FRB's host galaxy to lie behind our proposed host galaxy association. In this alternative scenario, the host galaxy would lie at a redshift $z_{\rm h} > 0.241$ that depends on the foreground galaxy's contribution to the total DM budget.
The FRB would pass through the putative intervening galaxy at a redshift $z_\mathcal{l} = 0.241$ at an impact parameter of about 4 kpc (based on the observed offset of the FRB localization in the optical images). The scattering contribution of the intervening galaxy lens is related to its DM contribution by\cite{Ocker2021, 2021arXiv210801172C}
\begin{equation}\label{eq:tauDM}
 \tau(\DM,\nu, z) \approx 48.03~{\rm \mu s} \times
 \frac{ \tilde{F} G_{\rm scatt}\, \DM_{\mathcal{l}}^2 }{(1 + z_{\mathcal{l}})^3 \nu^4},
\end{equation}
where $\DM_{\mathcal{l}}$ is the contribution of the intervening galaxy in \dmu~in its rest frame, $\nu$ is the observing frequency in GHz, $z_{\mathcal{l}}$ is the intervening galaxy redshift, and $\tilde{F} = \zeta \epsilon^2/f(l_{\rm o}^2 l_{\rm i})^{1/3}$ quantifies the electron density fluctuations in the scattering layer for $\tilde{F}$ in units of pc$^{-2/3}$ km$^{-1/3}$, where $\zeta$, $\epsilon^2$, and $f$ describe the density fluctuation statistics and filling factor, $l_{\rm o}$ is the outer scale of turbulence and $l_{\rm i}$ is the inner scale. The fluctuation parameter $\tilde{F}$ in the Milky Way varies from $\sim 10^{-3}$ pc$^{-2/3}$ km$^{-1/3}$ in the thick disk to $\gtrsim10^2$ pc$^{-2/3}$ km$^{-1/3}$ near the inner Galaxy, and is typically $\sim 0.1 - 1$ pc$^{-2/3}$ km$^{-1/3}$ in the thin disk, but it is not generally known for other galaxies \cite{Ocker2021}. The geometric factor $G_{\rm scatt}$ is unity for scattering in a host galaxy but it can be very large for an intervening galaxy, in which case $G_{\rm scatt} \approx 2d_{\rm sl}d_{\rm lo}/d_{\rm so}L$, where $d_{\rm sl}$, $d_{\rm lo}$, and $d_{\rm so}$ are the angular diameter distances between the source and lens, lens and observer, and source and observer, respectively, and $L$ is the path length through the lens. The numerical pre-factor in Equation~\ref{eq:tauDM} is for all angular diameter distances in Gpc and $L$ in Mpc.

\indent For the nominal DM contribution implied by the H$\alpha$ emission, ${\rm DM}_{\mathcal{l}} \approx 392$ pc cm$^{-3}$, and a path length through the intervening galaxy $L\approx 5$ kpc, the implied scattering time is $\tau \approx 20$ s at 1.25 GHz, orders of magnitude larger than the observed mean scattering time. For a smaller gas temperature $T_4 \sim 0.5$, yielding a DM contribution ${\rm DM}_{\mathcal{l}} \approx 290$ pc cm$^{-3}$, the implied scattering time is 
still $\tau \approx 18$ s at 1.25 GHz, assuming $\tilde{F} \sim 0.1$ pc$^{-2/3}$ km$^{-1/3}$ and $L \approx 5$ kpc. However, there is significant latitude in this estimate, depending on the assumed gas temperature, path length sampled by the FRB, and the value of $\tilde{F}$. While we have assumed that the FRB traces the entire ${\rm DM}_{\mathcal{l}}$ implied by the H$\alpha$ emission, the FRB LOS likely only traces a fraction of the H$\alpha$ emitting gas. If ${\rm DM}_{\mathcal{l}}$ is as small as 50 pc cm$^{-3}$, then $\tau$ could be as small as $0.6$ s for $\tilde{F} \sim 0.1$ pc$^{-2/3}$ km$^{-1/3}$ and $L \approx 5$ kpc.
While the implied scattering time could be two orders of magnitude smaller or larger depending on the combination of $\tilde{F}$, the path length $L$, and the parameters used to estimate the H$\alpha$ DM, our fiducial estimates suggest that the observed scattering time is significantly smaller than the scattering expected from an intervening galaxy. 

\subsection{Rotation measure search from FAST observations}

A search for rotation measure was performed with the FAST data.
The polarization was calibrated by correcting for differential gains and phases between the receptors through separate measurements of a noise diode injected at an angle of $45^{\circ}$ from the linear receptors. We searched for the RM from $-3.0\times10^5$ to $3.0\times10^5$\,$\mathrm{rad\,m^{-2}}$.
No significant peak was found in the Faraday spectrum. The observed lack of polarization could be due to the intrinsic low linear polarization, a depolarization process within the source or from intra-channel Faraday rotation. 

\end{methods}

\clearpage

\begin{addendum}

 \item[Acknowledgements]This work was supported by the Cultivation Project for FAST Scientific Payoff and Research Achievement of CAMS-CAS and also supported by National Natural Science Foundation of China (NSFC) Programs No. 11988101, No. 11725313, No. 11690024, No.12041303 No. U1731238, No. U2031117, No. U1831131, No. U1831207; CHN acknowledges support from the FAST Fellowship. 
 SC, JMC, and SO acknowledge support from the National Science Foundation (AAG~1815242) and are members of the NANOGrav Physics Frontiers Center, which is supported by NSF award PHY-1430284.
 CJL acknowledges support from the National Science Foundation under Grant No.~2022546. KA, SBS and RAT acknowledge support from NSF grant AAG-1714897. PW thanks the YSBR-006, CAS Project for Young Scientists in Basic Research. SBS is a CIFAR Azrieli Global Scholar in the Gravity and the Extreme Universe program. 
 YN acknowledges support from JSPS KAKENHI Grant Number JP20H01942. JJ acknowledges support from the Japan Society for the Promotion of Science (JSPS) KAKENHI grant JP19K23456 and JP18J12714. 
 SD is the recipient of an Australian Research Council Discovery Early Career Award (DE210101738) funded by the Australian Government.
 We thank the FAST key science project for supporting follow-up observations. We also thank the FAST collaborations and realfast team for their technical support.
 Some data presented herein were obtained at the W. M. Keck Observatory, which is operated as a scientific partnership among the California Institute of Technology, the University of California and the National Aeronautics and Space Administration. The Observatory was made possible by the generous financial support of the W. M. Keck Foundation.
 This study is based in part on data collected at the Subaru Telescope, which is operated by the National Astronomical Observatory of Japan. The National Radio Astronomy Observatory is a facility of the National Science Foundation operated under cooperative agreement by Associated Universities, Inc. This work was supported by China Science and Technology Cloud(CSTCloud) and China Environment for Network Innovations (CENI). We thank the staff of CSTCloud/CENI for their support during data processing.
 
\item[Author contributions] CHN discovered the source \frb. DL, CL, SC, CWT, WY, and CHN initiated the follow-up projects. DL, CHN, BZ, and WWZ led the follow-up FAST observations. CHN, JMY, YKZ, PW, DJZ, and YF searched and processed the FAST data.
KA, CL, SC, XZ, SBS, ZY, WY, and LC contributed to the VLA burst detection and localization, identification and measurements of the associated persistent radio source. CWT, SC, DS, YN, JJ, CB, and GDL contributed to the optical/NIR follow-up observations and analysis.
SKO, JMC, SC, JMY, and CHN measured the burst scattering and analysed the propagation effects. JMC, BZ, and WYW contributed to the $\rm DM_{host}$ estimation.
KA, YKZ, JRN, RL, WWZ, and CHN contributed to the periodicity and burst rate analysis. PW and YKZ helped with energy calibration and MY contributed to the radio frequency interference removal on FAST data. ZY, WY, MC, SD, and YLY contributed to other follow-up observations.
SBS, DL, KA, CHN, SC, JMC, SKO, and CL had major contributions to the preparation of the manuscript. All of the authors have reviewed, discussed, and commented on the presented results and the manuscript. 

\item[Supplementary Information] is available for this paper.

 \item[Competing interests] The authors declare that they have no competing financial interests. 
 
 \item[Correspondence and requests for materials] should be addressed to D. Li (dili@nao.cas.cn), W. Yu (wenfei@shao.ac.cn), C. J. Law    (claw@astro.caltech.edu)
 
 
 
 \item[Reprints and permissions information is available at \url{www.nature.com/reprints}.]
 
 \end{addendum}

\clearpage

\section*{Extended Data Legends}

\begin{enumerate}[label=\textbf{ED Figure \arabic*},ref=ED Figure \arabic*, wide]
\item \textbf{$~|~$Positions of the bursts and persistent radio source identified with VLA observations.} The observation were performed at 1.5, 3, and 5.5~GHz. They were shown as offsets from the best-fit position of the ensemble of bursts, at R.A. = 16h02m04.272s, Decl. = $-11$\arcdeg17\arcmin17.32\arcsec (J2000). The uncertainties on the positions of the bursts are indicated with shaded ellipses, and those for the PRS are shown with error bars. These uncertainties include $1\sigma$ statistical errors and estimates for systematic errors added in quadrature. \label{fig:astrometry} 

\item \textbf{$~|~$Spectrum of the PRS associated with \frb.} The bandwidth is split into two subbands for all the observations and the corresponding radio flux densities are shown. The observation dates in MJD are shown in the legend. The frequencies for each subband are shifted slightly in order to show the flux density error bars. The inset plot shows the average flux at each subband. A power-law model fit to these measurements is shown by the black dashed line and yields a spectral index of $-0.41~\pm$ 0.04 for the PRS. \label{fig:spectrum}

\item \textbf{$~|~$Optical spectrum of the \frb\ host galaxy obtained at Palomar.} The redshift of $z = 0.241$ was determined using the [OIII]-5007\AA\ line and H$\alpha$ line ($> 5 \sigma$ detections). These two emission lines are narrow, with FWHM $\sim 10$~\AA. The flux scale of the spectrum is not corrected for slit loss or extinction. \label{fig:spectrum_line}

\item \textbf{$~|~$Posterior probability distributions.} The probability distribution are for the shape parameter $k$ and event rate $r$ of the Weibull distribution. a) Results from a fit to all burst waiting times of \frb~detected by FAST. b) Results from a fit to all burst waiting times longer than 1~s. These bursts are at fluences higher than 9 mJy ms.\label{fig:weibull}

\item \textbf{$~|~$The fluence-width distribution at 1.25 GHz for \frb.} The widths and fluences of the 79 bursts detected by FAST are also shown in supplementary information Table 1.\label{fig:fluence}

\item \textbf{$~|~$The distribution of burst energies for \frb.} The bursts are from FAST observations. The red line and black dashed line are the log-normal fit and 90$\%$ completeness threshold, respectively. The 90$\%$ completeness threshold uses 0.023 Jy~ms, which is the simulation result from ref.~\cite{2021Natur.598..267L}.\label{fig:enhist}

\item \textbf{$~|~$Dynamic spectra of bursts with and without scattering.} a) Frequency-time dynamic spectra of bursts with significant evidence of pulse broadening, along with 1D burst profiles that have been averaged across the entire frequency band. b) Examples of bursts without significant evidence of pulse broadening. White lines indicate excised radio frequency interference. All 1D burst profiles are shown in units of the signal-to-noise ratio (S/N). For P31, there is a potential combination of scattering, time-frequency drift, and multiple unresolved burst components; in this case we report a scattering time that should be interpreted as an upper limit. Bursts with scattering time constraints are shown in supplementary information Table 1. \label{fig:scattering_examples}

\end{enumerate}

\begin{enumerate}[label=\textsc{\bf{ED Table \arabic*}},ref=ED Table \arabic*, wide]


\item  \textbf{$~|~$VLA observations of the PRS and number of bursts detected in each observation.} Frequency refers to the center of the observing band and duration represents the total length of on-source observations. Note that the \realf\ system was not run on the last observation due to a system error. \label{tab:vla_observations}

\item \textbf{$~|~$Localised positions of the PRS from 1.5, 3, 5.5 GHz VLA deep images.} The first column shows the observing central frequency. The second and third columns report the coordinates of the PRS from the deep images. The remaining columns show the 1 $\sigma$ statistical errors and the cross-matched systematic offsets in right ascension and declination. \label{tab:vla_prs_locs}

\item \textbf{$~|~$Localised positions of the VLA bursts.} The label in the first column shows the frequency band of observation, followed by the burst number. DM represents the S/N maximizing DM obtained using offline refinement of the bursts and S/N reports the maximum image S/N. The remaining columns show the burst positions (and errors) obtained using CASA calibration and image fitting. The systematic errors reported here were estimated using deep radio images generated using all the observations at the respective frequency band. \label{tab:vla_burst_locs}

\item \textbf{$~|~$Properties of the VLA bursts.} The label in the first column shows the frequency band of observation, followed by the burst number. MJD is referenced to the solar system barycenter (in the TDB scale) and corrected for dispersion delay at infinite frequency. Width values here should be considered as upper limits. Frequency represents the range of frequencies in which the burst signal is prominently seen, and was visually determined. The DM that maximizes the S/N of the respective burst is given in the third column. The apparent DMs may reflect time-frequency structure, as opposed to bona fide DM variations. \label{tab:vla_burst_props}

\end{enumerate}





\clearpage
\newpage
\renewcommand{\thefigure}{Figure \arabic{figure} } 
\renewcommand{\thetable}{Table \arabic{table} } 
\renewcommand{\figurename}{\textbf{Extended Data $|$ }}
\renewcommand{\tablename}{\textbf{Extended Data $|$ }}

\section*{Extended Data Legends}

\begin{figure*}[h!]
\includegraphics[width=0.7\textwidth]{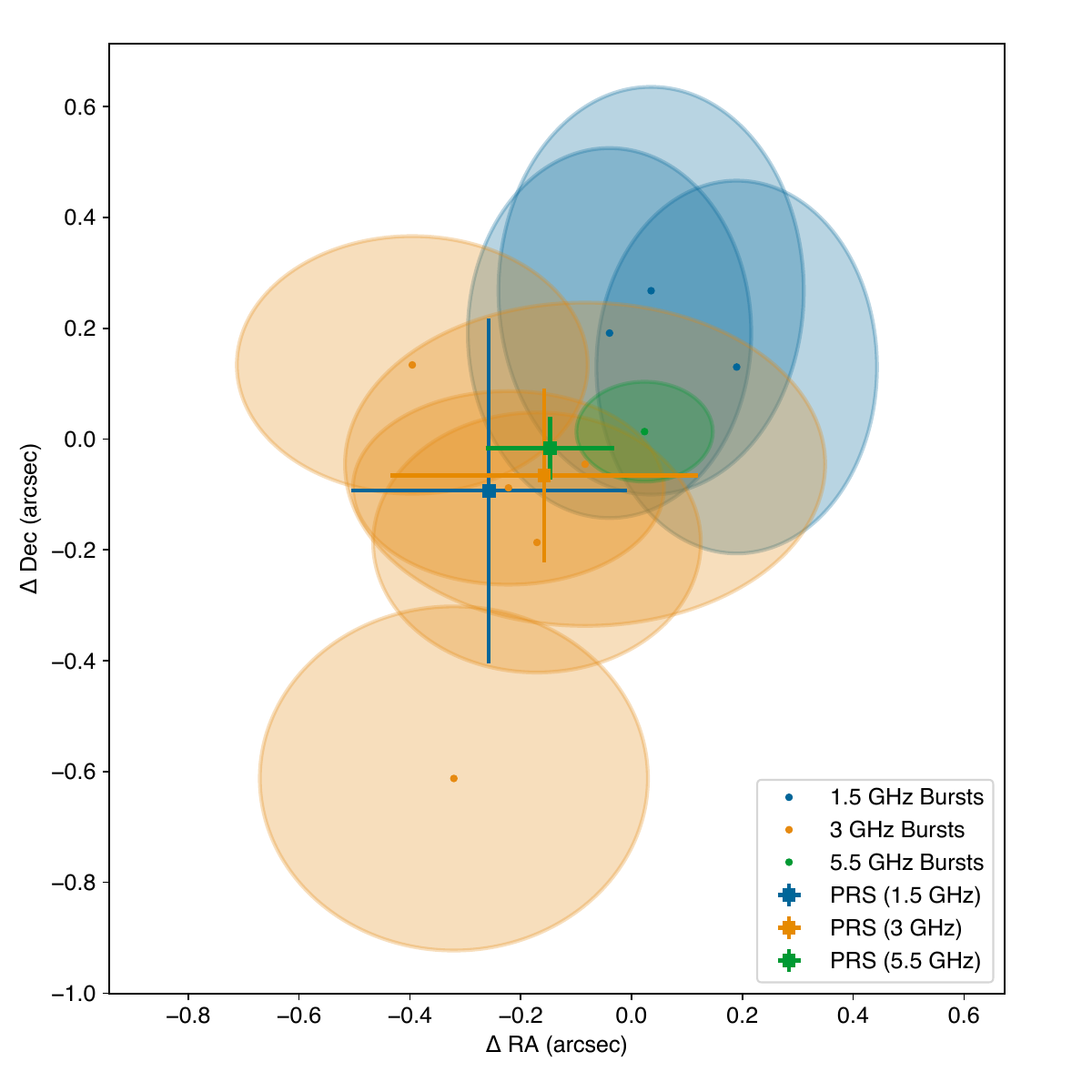}
\caption{\footnotesize \textbf{Positions of the bursts and persistent radio source identified with VLA observations.} The observation were performed at 1.5, 3, and 5.5~GHz. They were shown as offsets from the best-fit position of the ensemble of bursts, at R.A. = 16h02m04.272s, Decl. = $-11$\arcdeg17\arcmin17.32\arcsec (J2000). The uncertainties on the positions of the bursts are indicated with shaded ellipses, and those for the PRS are shown with error bars. These uncertainties include $1\sigma$ statistical errors and estimates for systematic errors added in quadrature.}
\end{figure*}

\begin{figure*}
\centering
\includegraphics[width=0.8\textwidth]{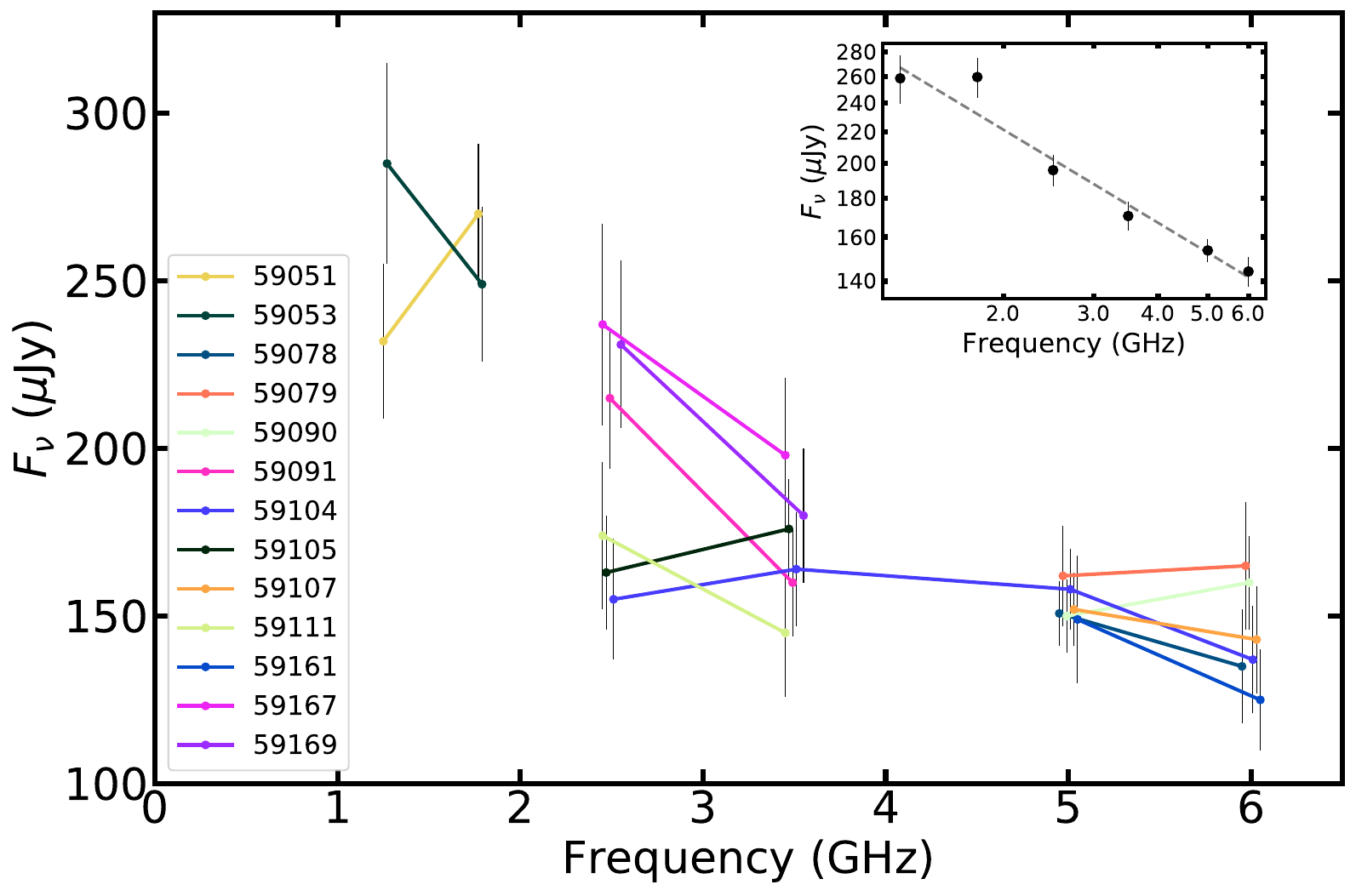}
\caption{\footnotesize \textbf{Spectrum of the PRS associated with \frb.} The bandwidth is split into two subbands for all the observations and the corresponding radio flux densities are shown. The observation dates in MJD are shown in the legend. The frequencies for each subband are shifted slightly in order to show the flux density error bars. The inset plot shows the average flux at each subband. A power-law model fit to these measurements is shown by the black dashed line and yields a spectral index of $-0.41~\pm$ 0.04 for the PRS.} 
\end{figure*}

\begin{figure*}
\centering
\includegraphics[width=0.8\textwidth]{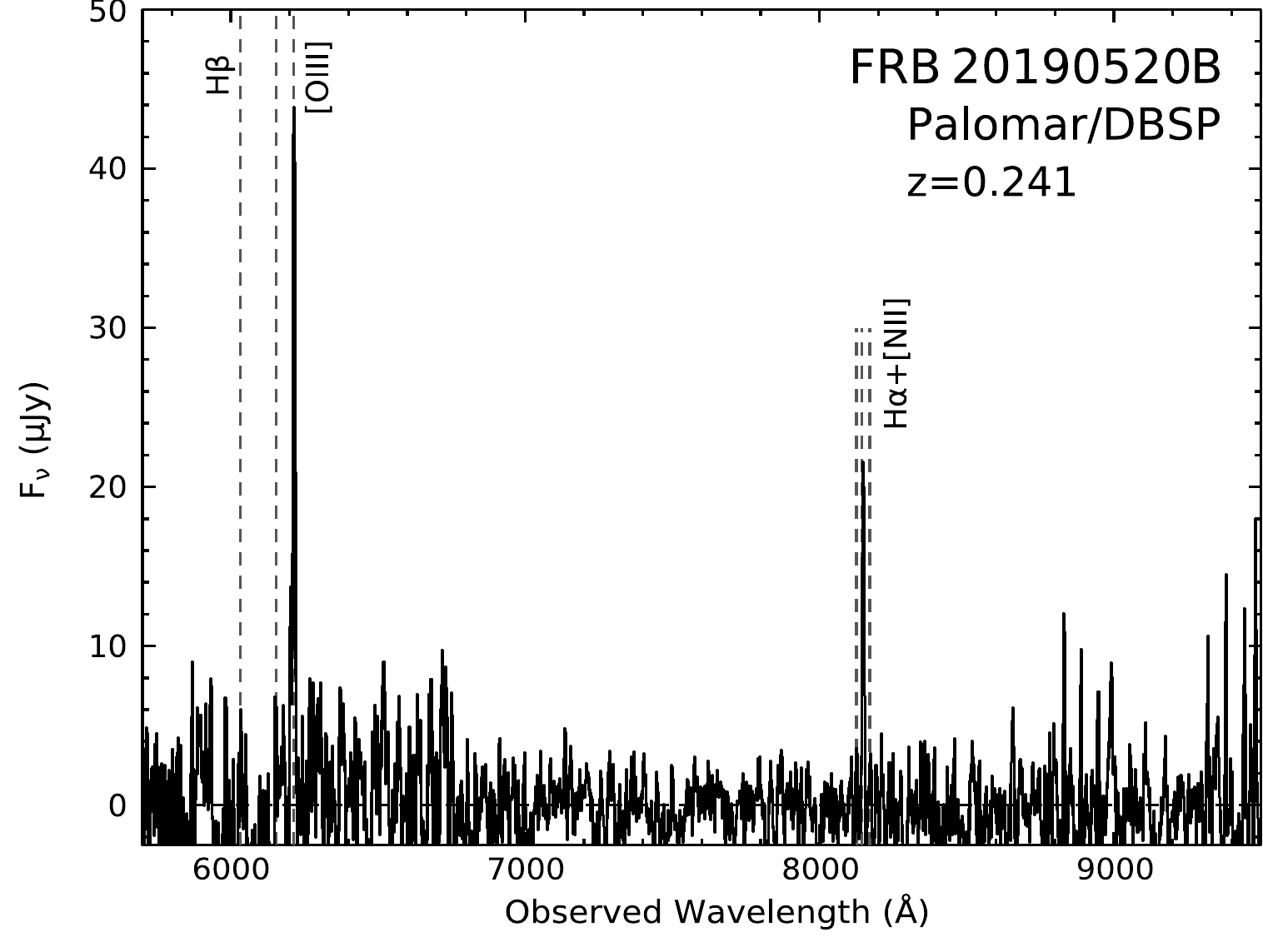}
\caption{\footnotesize\textbf{Optical spectrum of the \frb\ host galaxy obtained at Palomar.} The redshift of $z = 0.241$ was determined using the [OIII]-5007\AA\ line and H$\alpha$ line ($> 5 \sigma$ detections). These two emission lines are narrow, with FWHM $\sim 10$~\AA. The flux scale of the spectrum is not corrected for slit loss or extinction. }
\end{figure*}

\begin{figure*}
\centering
\includegraphics[width=0.9\textwidth]{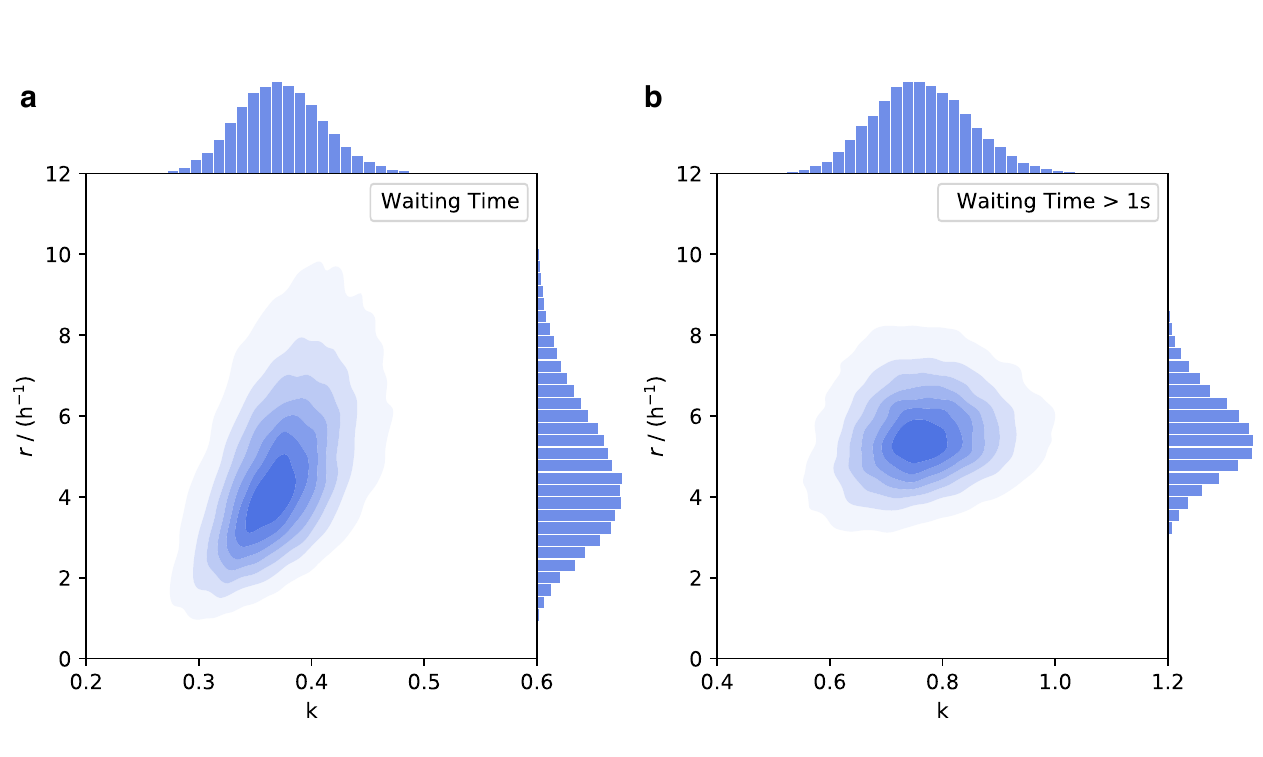}
\caption{\footnotesize \textbf{Posterior probability distributions.} The probability distribution are for the shape parameter $k$ and event rate $r$ of the Weibull distribution. a) Results from a fit to all burst waiting times of \frb~detected by FAST. b) Results from a fit to all burst waiting times longer than 1~s. These bursts are at fluences higher than 9 mJy ms.}
\end{figure*}

\begin{figure*}
\centering
\includegraphics[width=0.9\textwidth]{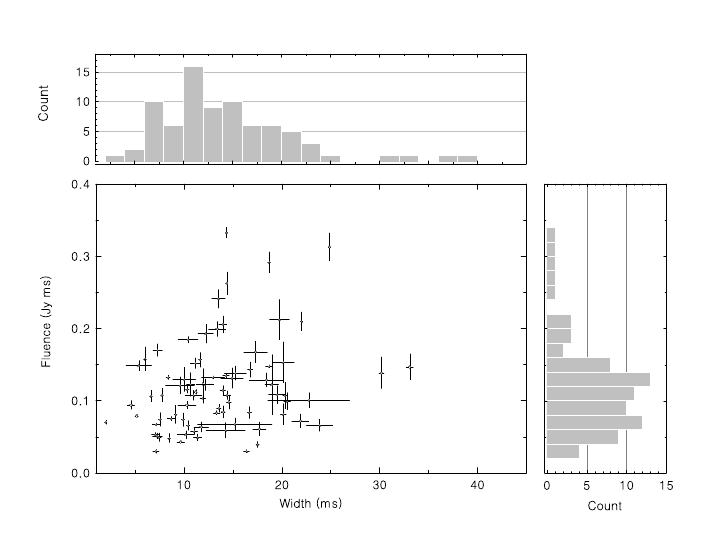}
\caption{\footnotesize\textbf{The fluence-width distribution at 1.25 GHz for \frb.} The widths and fluences of the 79 bursts detected by FAST are also shown in supplementary information Table 1.} 
\end{figure*}

\begin{figure*}
\centering
\includegraphics[width=0.9\textwidth]{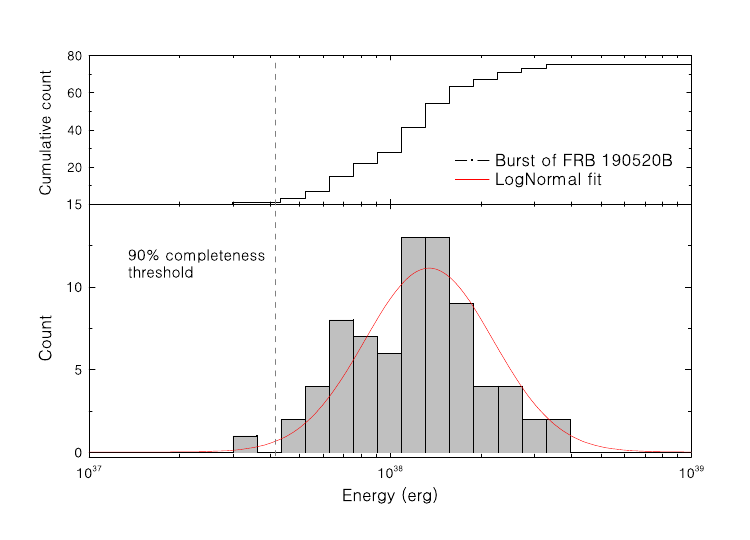}
\caption{\footnotesize\textbf{The distribution of burst energies for \frb.} The bursts are from FAST observations. The red line and black dashed line are the log-normal fit and 90$\%$ completeness threshold, respectively. The 90$\%$ completeness threshold is 0.023 Jy~ms.}
\end{figure*}

\begin{figure*}[ht]
\centering
\includegraphics[width=0.8\textwidth]{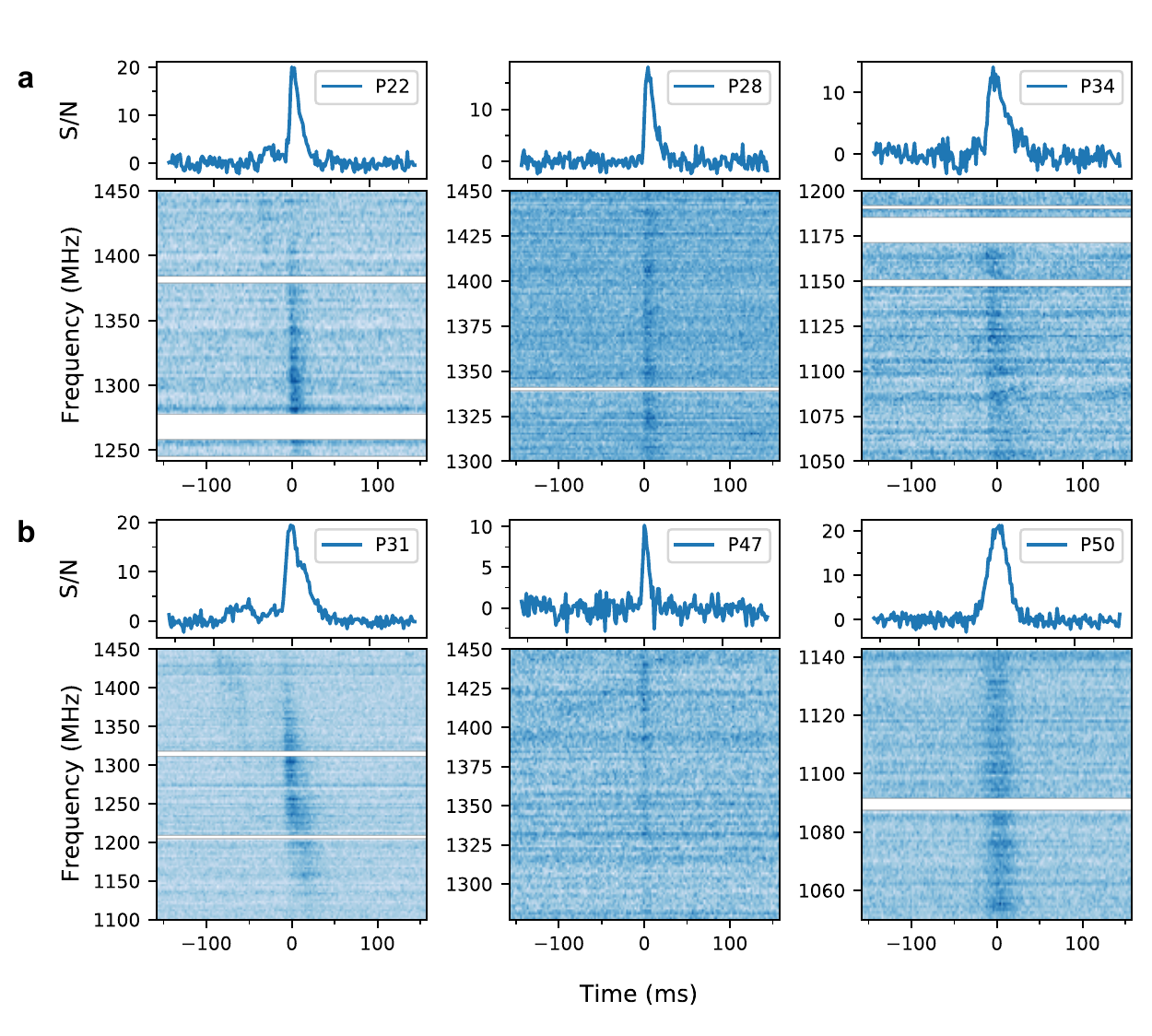}
\caption{\footnotesize\textbf{Dynamic spectra of bursts with and without scattering.} a) Frequency-time dynamic spectra of bursts with significant evidence of pulse broadening, along with 1D burst profiles that have been averaged across the entire frequency band. b) Examples of bursts without significant evidence of pulse broadening. White lines indicate excised radio frequency interference. All 1D burst profiles are shown in units of the signal-to-noise ratio (S/N). For P31, there is a potential combination of scattering, time-frequency drift, and multiple unresolved burst components; in this case we report a scattering time that should be interpreted as an upper limit. Bursts with scattering time constraints are shown in supplementary information Table 1. }
\end{figure*}

\clearpage
\renewcommand\arraystretch{0.8}
\scriptsize

\begin{table}
\rmfamily
\centering
\footnotesize

\begin{tabular}{|l|c|c|c|c|c|}
\hline
Start Time & Frequency & Duration & Beam size & Flux density & Bursts\\
(MJD) & (GHz) & (min) & ($^{\prime\prime}$,$^{\prime\prime}$) & ($\mu$Jy) & \\
\hline 
59051.033815 & 1.5 & 96 & 4.78$\times$2.90 & 258$\pm$16 & 1 \\
59053.046790 & 1.5 & 96 & 4.63$\times$2.90 & 273$\pm$23 & 2 \\
59079.004955 & 5.5 & 41 & 1.38$\times$1.02 & 145$\pm$10 & 0 \\
59079.971533 & 5.5 & 41 & 1.44$\times$1.05 & 164$\pm$11 & 0 \\
59090.941498 & 5.5 & 41 & 1.45$\times$1.04 & 158$\pm$9 & 0 \\
59091.938768 & 3 & 41 & 2.43$\times$1.85 & 195$\pm$14 & 5 \\
59104.867225 & 3 & 41 & 2.76$\times$1.76 & 160$\pm$13 & 0 \\
59104.908789 & 5.5 & 41 & 1.43$\times$1.04 & 151$\pm$9 & 0 \\
59105.976837 & 3 & 41 & 2.44$\times$1.60 & 186$\pm$15 & 0 \\
59107.915546 & 5.5 & 41 & 1.38$\times$1.02 & 153$\pm$10 & 1 \\
59111.116215 & 3 & 41 & 5.00$\times$1.45 & 176$\pm$16 & 0 \\
59161.677240 & 5.5 & 41 & 1.92$\times$0.47 & 139$\pm$13 & 0 \\
59167.637761 & 3 & 41 & 2.76$\times$0.51 & 233$\pm$17 & 0 \\
59169.640486 & 3 & 41 & 2.47$\times$0.48 & 211$\pm$14 & - \\ 
\hline 
\end{tabular} 

\caption{\textbf{VLA observations of the PRS and number of bursts detected in each observation.} Frequency refers to the center of the observing band and duration represents the total length of on-source observations. Note that the \realf\ system was not run on the last observation due to a system error.} 
\end{table}

\begin{table}
\rmfamily
\centering
\footnotesize
\begin{tabular}{ccccccc}
\hline 
Frequency & R.A. & Decl. & 
\multicolumn{2}{c}{Statistical Error} & \multicolumn{2}{c}{Systematic Offsets} \\ 
(GHz)& & & R.A. ($^{\prime\prime}$) & Decl. ($^{\prime\prime}$) & R.A. ($^{\prime\prime}$) & Decl. ($^{\prime\prime}$) \\ 
\hline 
1.5& 16h02m04.2543s & $-11$d17m17.4146s & 0.0386 & 0.0713 & 0.0373$\pm$0.2460 & 0.0144$\pm$0.3035 \\
3& 16h02m04.2611s & $-11$d17m17.3869s & 0.0176 & 0.0168 & $-$0.0696$\pm$0.2772 & 0.0382$\pm$0.1552 \\
5.5& 16h02m04.2618s & $-11$d17m17.3375s & 0.0075 & 0.0095 & $-$0.0947$\pm$0.1161 & 0.1069$\pm$0.0554 \\
\hline 
\end{tabular}
\caption{\textbf{Localised positions of the PRS from 1.5, 3, 5.5 GHz VLA deep images.} The first column shows the observing central frequency. The second and third columns report the coordinates of the PRS from the deep images. The remaining columns show the 1 $\sigma$ statistical errors and the cross-matched systematic offsets in right ascension and declination.}
\end{table} 

\begin{table}
\rmfamily
\centering
\footnotesize
\begin{tabular}{cccccccc}
\hline 
Band & S/N & R.A. & Decl. & \multicolumn{2}{c}{Statistical Error} & \multicolumn{2}{c}{Systematic Error} \\ 
& & & & R.A. ($^{\prime\prime}$) & Decl. ($^{\prime\prime}$) & R.A. ($^{\prime\prime}$) & Decl. ($^{\prime\prime}$)\\ 
\hline 
L1 & 16 & 16h02m04.2742s & $-11$d17m17.0535s & 0.124 & 0.208 & 0.2460 & 0.3035 \\
L2 & 27 & 16h02m04.2847s & $-11$d17m17.1912s & 0.060 & 0.145 & 0.2460 & 0.3035 \\
L3 & 15 & 16h02m04.2691s & $-11$d17m17.1299s & 0.070 & 0.139 & 0.2460 & 0.3035 \\
S1 & 19 & 16h02m04.2567s & $-11$d17m17.4093s & 0.044 & 0.080 & 0.2772 & 0.1552 \\
S2 & 17 & 16h02m04.2449s & $-11$d17m17.1874s & 0.153 & 0.173 & 0.2772 & 0.1552 \\
S3 & 15 & 16h02m04.2602s & $-11$d17m17.5078s & 0.103 & 0.176 & 0.2772 & 0.1552 \\
S4 & 10 & 16h02m04.2661s & $-11$d17m17.3666s & 0.333 & 0.247 & 0.2772 & 0.1552 \\
S5 & 7  & 16h02m04.2500s & $-11$d17m17.9337s & 0.214 & 0.269 & 0.2772 & 0.1552 \\
C1 & 19 & 16h02m04.2734s & $-11$d17m17.3078s & 0.041 & 0.071 & 0.1161 & 0.0554 \\
\hline 
\end{tabular}
\caption{\textbf{Localised positions of the VLA bursts.} The label in the first column shows the frequency band of observation, followed by the burst number. S/N reports the maximum image S/N. The remaining columns show the burst positions (and errors) obtained using CASA calibration and image fitting. The systematic errors reported here were estimated using deep radio images generated using all the observations at the respective frequency band.}
\end{table}

\begin{table}
\rmfamily
\centering
\footnotesize
\begin{tabular}{cccccccc}
\hline 
Name & MJD & DM & Flux density (Jy) & Width (ms) & Frequency (GHz)\\
\hline
L1 & 59051.0476715 & 1209.60 & 0.113$\pm$0.006 & 20 & 1.33--1.52 \\
L2 & 59053.1045803 & 1227.15 & 0.275$\pm$0.010 & 20 & 1.39--1.52 \\
L3 & 59053.0973272 & 1236.60 & 0.189$\pm$0.009 & 10 & 1.78--1.97 \\
S1 & 59091.9604318 & 1291.30 & 0.075$\pm$0.004 & 20 & 3.12--3.50 \\
S2 & 59091.9560792 & 1193.40 & 0.100$\pm$0.009 & 20 & 2.63--2.88 \\
S3 & 59091.9635008 & 1222.75 & 0.093$\pm$0.007 & 20 & 2.50--2.76 \\
S4 & 59091.9435287 & 1216.80 & 0.037$\pm$0.005 & 10 & 2.88--3.24 \\
S5 & 59091.9668476 & 1276.50 & 0.022$\pm$0.003 & 10 & 2.50--3.50 \\ 
C1 & 59107.9272138 & 1267.50 & 0.068$\pm$0.004 & 10 & 4.49--4.88 \\
\hline 
\end{tabular}
\caption{\textbf{Properties of the VLA bursts.} The label in the first column shows the frequency band of observation, followed by the burst number. MJD is referenced to the solar system barycenter (in the TDB scale) and corrected for dispersion delay at infinite frequency. Width values here should be considered as upper limits. Frequency represents the range of frequencies in which the burst signal is prominently seen, and was visually determined. The DM that maximizes the S/N of the respective burst is given in the third column. The apparent DMs may reflect time-frequency structure, as opposed to bona fide DM variations.}
\end{table} 

\clearpage

\newpage
\setcounter{figure}{0}
\setcounter{table}{0}
\renewcommand{\thetable}{Table \arabic{table} } 
\renewcommand{\tablename}{\textbf{Supplementary Information $|$ }}
\footnotesize

\setlength{\tabcolsep}{2mm}{
\setlength{\LTcapwidth}{7in}
\renewcommand\arraystretch{0.7}

\begin{longtable}{c c c c c c c} 
\caption{\centering\textbf{Properties of the 79 \frb~ bursts detected by FAST.}}\\
\hline
\hline
Burst ID$^{a}$ & Burst time$^{b}$ & DM$^{c}$ & Pulse width & Scatter tail$^{d}$ & Fluence & Energy$^{e}$\\
& (MJD) & (\dmu) & (ms)  & (ms) & (mJy\ ms) & ($\times10^{37}$erg) \\
\hline
\hline
\endhead
\hline
\hline
\endfoot
\hline
\hline
\endlastfoot
 \textbf{P1} & 58623.716958197 & 1238.6(20)$^{f}$ & 8.7(8) & - & 75(3) & 9.1(4) \\
 P2 & 58623.716997034 & 1238.6(20) & 16.4(6) & - & 30(2) & 3.6(2) \\
 P3 & 58623.717075743 & 1238.6(20) & 7.1(8) & - & 53(3) & 6.4(4) \\
 P4 & 58623.717421629 & 1238.6(20) & 7.3(10) & - & 51(4) & 6.2(5) \\
\hline \textbf{P5} & 58963.766862939 & 1209.1(10) & 10.4(2) & - & 65(6) & 8.0(8) \\
 P6 & 58963.790991383 & 1209.1(10) & 10.2(17) & 13.1(23) & 52(5) & 6.4(6) \\
\hline P7 & 58991.684351079 & 1214.2(5) & 20.3(4) & - & 107(18) & 13.0(22) \\
 P8 & 58991.686137870 & 1214.2(5) & 22.8(83) & - & 100(10) & 12.2(12) \\
 P9 & 58991.686875951 & 1214.2(5) & 9.2(2) & - & 81(12) & 9.8(15) \\
 P10 & 58991.704636888 & 1214.2(5) & 19.5(17) & 3.6(8) & 108(12) & 13.1(15) \\
 P11 & 58991.704638485 & 1214.2(5) & 9.7(7) & - & 42(0) & 5.2(0) \\
 P12 & 58991.717697335 & 1214.2(5) & 11.9(11) & 10.3(12) & 122(21) & 14.9(25) \\
 P13 & 58991.717878516 & 1214.2(5) & 14.3(40) & 5.1(25) & 58(10) & 7.1(12) \\
 P14 & 58991.718217265 & 1214.2(5) & 11.8(15) & - & 63(6) & 7.7(8) \\
 P15 & 58991.718217670 & 1214.2(5) & 17.5(3) & - & 39(4) & 4.8(5) \\
 \textbf{P16} & 58991.718218082 & 1214.2(5) & 20.2(22) & 12.6(19) & 153(28) & 18.6(34) \\
 P17 & 58991.735450006 & 1214.2(5) & 30.2(5) & - & 138(21) & 16.8(26) \\
 P18 & 58991.750640038 & 1214.2(5) & 17.7(13) & - & 60(9) & 7.4(11) \\
 P19 & 58991.750640790 & 1214.2(5) & 20.2(5) & - & 81(14) & 9.9(17) \\
\hline P20 & 59060.484475154 & 1209.1(5) & 13.5(13) & 11.3(14) & 241(12) & 29.3(15) \\
 P21 & 59060.507858658 & 1209.1(5) & 7.3(9) & 9.3(12) & 170(8) & 20.7(10) \\
 \textbf{P22} & 59060.525960600 & 1209.1(5) & 17.3(24) & 7.6(16) & 167(14) & 20.3(17) \\
\hline P23 & 59061.512755579 & 1209.1(6) & 8.5(1) & - & 48(6) & 5.9(7) \\
 \textbf{P24} & 59061.512755780 & 1209.1(6) & 10.0(23) & 4.7(17) & 129(16) & 15.7(20) \\
 P25 & 59061.516277966 & 1209.1(6) & 7.1(6) & - & 29(2) & 3.6(3) \\
 P26 & 59061.516278904 & 1209.1(6) & 7.5(4) & - & 49(5) & 6.0(7) \\
 P27 & 59061.516279436 & 1209.1(6) & 11.4(8) & - & 49(4) & 6.0(5) \\
 P28 & 59061.524341261 & 1209.1(6) & 10.7(8) & 10.4(9) & 122(16) & 14.8(19) \\
 P29 & 59061.535633328 & 1209.1(6) & 14.0(4) & - & 84(8) & 10.2(10) \\
 P30 & 59061.536568858 & 1209.1(6) & 21.9(17) & - & 72(9) & 8.7(11) \\
 P31 & 59061.536569628 & 1209.1(6) & 19.0(14) & 14.0(13) & 122(41) & 14.8(50) \\
 P32 & 59061.537903820 & 1209.1(6) & 23.9(27) & 4.9(13) & 65(8) & 8.0(9) \\
 P33 & 59061.539298603 & 1209.1(6) & 9.9(3) & - & 73(9) & 8.9(11) \\
 P34 & 59061.541828100 & 1209.1(6) & 19.8(20) & 8.7(10) & 212(28) & 25.8(34) \\
\hline P35 & 59067.467544345 & 1202.8(6) & 15.2(76) & - & 67(8) & 8.2(10) \\
 P36 & 59067.486738799 & 1202.8(6) & 14.9(22) & 17.8(29) & 131(14) & 15.9(17) \\
 \textbf{P37} & 59067.486739378 & 1202.8(6) & 33.1(7) & - & 146(18) & 17.8(21) \\
 P38 & 59067.502691880 & 1202.8(6) & 14.3(6) & - & 134(3) & 16.3(3) \\
 P39 & 59067.509899127 & 1202.8(6) & 12.2(18) & 10.3(17) & 122(8) & 14.8(9) \\
 P40 & 59067.535246460 & 1202.8(6) & 7.6(2) & - & 74(9) & 9.0(11) \\
\hline P41 & 59069.495909561 & 1190.2(11) & 16.8(5) & - & 143(10) & 17.4(12) \\
 \textbf{P42} & 59069.501196109 & 1190.2(11) & 18.7(1) & - & 291(14) & 35.4(17) \\
 P43 & 59069.514994796 & 1190.2(11) & 12.2(15) & 11.6(17) & 192(13) & 23.4(15) \\
\hline P44 & 59071.472522775 & 1200.0(11) & 5.2(3) & - & 79(1) & 9.6(2) \\
 P45 & 59071.472523007 & 1200.0(11) & 7.2(7) & - & 66(1) & 8.1(2) \\
 \textbf{P46} & 59071.491696655 & 1200.0(11) & 10.4(20) & 4.9(15) & 184(4) & 22.4(5) \\
\hline P47 & 59073.496887082 & 1197.0(7) & 7.8(2) & - & 107(9) & 13.1(11) \\
 \textbf{P48} & 59073.515256071 & 1197.0(7) & 18.7(6) & - & 147(2) & 17.9(2) \\
 P49 & 59073.515256881 & 1197.0(7) & 11.1(7) & - & 57(4) & 7.0(5) \\
\hline \textbf{P50} & 59075.454353002 & 1210.6(6) & 24.9(1) & - & 313(19) & 38.0(23) \\
 P51 & 59075.454862469 & 1210.6(6) & 22.0(1) & - & 209(12) & 25.5(15) \\
 P52 & 59075.472181012 & 1210.6(6) & 14.0(5) & - & 114(6) & 13.8(8) \\
 P53 & 59075.484186304 & 1210.6(6) & 6.0(0) & - & 157(16) & 19.1(20) \\
 P54 & 59075.496463775 & 1210.6(6) & 16.7(4) & - & 83(8) & 10.1(9) \\
\hline P55 & 59077.448938437 & 1209.9(4) & 11.9(5) & - & 103(7) & 12.5(8) \\
 P56 & 59077.448939131 & 1209.9(4) & 14.4(7) & - & 107(5) & 13.1(7) \\
 P57 & 59077.449538432 & 1209.9(4) & 13.4(17) & 10.6(16) & 199(10) & 24.2(12) \\
 P58 & 59077.449939184 & 1209.9(4) & 5.4(26) & - & 148(7) & 18.0(8) \\
 P59 & 59077.449939372 & 1209.9(4) & 9.6(29) & - & 121(11) & 14.7(14) \\
 P60 & 59077.460100338 & 1209.9(4) & 13.0(23) & 13.5(28) & 132(1) & 16.1(2) \\
 P61 & 59077.460503968 & 1209.9(4) & 10.3(18) & 10.6(21) & 94(5) & 11.4(6) \\
 P62 & 59077.466295203 & 1209.9(4) & 11.7(2) & - & 156(10) & 19.0(12) \\
 P63 & 59077.468715062 & 1209.9(4) & 18.4(35) & 7.6(18) & 128(9) & 15.6(11) \\
 P64 & 59077.469903193 & 1209.9(4) & 11.2(10) & 11.3(12) & 152(7) & 18.5(9) \\
 P65 & 59077.475047947 & 1209.9(4) & 11.0(17) & 8.3(16) & 107(6) & 13.0(8) \\
 P66 & 59077.475330491 & 1209.9(4) & 14.0(8) & 12.9(9) & 205(11) & 24.9(14) \\
 P67 & 59077.477447949 & 1209.9(4) & 2.0(1) & - & 70(2) & 8.5(3) \\
 \textbf{P68} & 59077.485451841 & 1209.9(4) & 14.3(1) & - & 332(7) & 40.4(8) \\
 P69 & 59077.485451957 & 1209.9(4) & 14.4(1) & - & 262(15) & 31.8(19) \\
 P70 & 59077.490027004 & 1209.9(4) & 8.4(4) & - & 132(2) & 16.1(3) \\
 P71 & 59077.491413589 & 1209.9(4) & 10.3(3) & - & 115(5) & 14.0(6) \\
 P72 & 59077.497806611 & 1209.9(4) & 15.3(22) & 10.7(21) & 138(9) & 16.8(11) \\
 P73 & 59077.497957805 & 1209.9(4) & 13.6(4) & - & 89(5) & 10.9(6) \\
 P74 & 59077.498960966 & 1209.9(4) & 4.6(7) & 11.8(15) & 94(6) & 11.4(7) \\
\hline \textbf{P75} & 59089.428163703 & 1186.7(25) & 11.2(2) & - & 112(3) & 13.6(4) \\
 P76 & 59089.436165435 & 1186.7(25) & 13.3(6) & - & 83(2) & 10.1(3) \\
\hline \textbf{P77} & 59111.370525959 & 1183.3(17) & 20.6(6) & - & 99(11) & 12.0(14) \\
 P78 & 59111.370526098 & 1183.3(17) & 6.7(1) & - & 106(8) & 12.9(9) \\
 P79 & 59111.370978062 & 1183.3(17) & 14.6(5) & - & 98(9) & 11.9(11) \\
\end{longtable}}

\begin{tablenotes}
\spacing{1.2}
\footnotesize
\item[a] $^a$ The burst IDs are from 1 to 79. P1 to P4 were detected in the drift scan mode. P4 to P79 were detected in the tracking mode.
\item[b] $^b$ Arrival time of burst at the solar system barycenter in barycentric dynamical time (TDB). They are corrected to the frequency of 1.5 GHz in the International Celestial Reference System (ICRS).
\item[c] $^c$ All the bursts detected on the same day are assigned the best fit DM value of the highest S/N burst from that day; the highest S/N burst ID is shown in bold for each observing epoch. Epochs are separated by single horizontal lines and the apparent DM variations may solely be the result of the variable spectra-temporal structure of the bursts.
\item[d] $^d$ We only report burst scattering timescales with fractional uncertainties less than 50$\%$. These scattering times are based on fitting a Gaussian pulse convolved with a one-sided exponential to the 1D burst profile, which was averaged over the entire burst bandwidth. 
\item[e] $^e$ Energy here refers to equivalent isotropic energy.
\item[f] $^f$ Reported uncertainties correspond to the last consecutive digits; e.g., $1238.6(20) = 1238.6 \pm 2.0$. 
\end{tablenotes}
\end{document}